\documentclass[12pt]{article}
\usepackage[margin=1.9cm]{geometry}
\usepackage{epsfig}
\usepackage{cite}

\newcommand{\mysection}{\setcounter{equation}{0}\section}

\def\beq{\begin{equation}}
\def\eeq{\end{equation}}
\def\beqa{\begin{eqnarray}}
\def\eeqa{\end{eqnarray}}
 
\begin{document}

\begin{center}
{\Large \bf QCD corrections in $tq\gamma$ production at hadron colliders}
\end{center}

\vspace{2mm}

\begin{center}
{\large Nikolaos Kidonakis$^a$ and Nodoka Yamanaka$^{b,c}$}\\

\vspace{2mm}

${}^a${\it Department of Physics, Kennesaw State University, \\
Kennesaw, GA 30144, USA}

\vspace{1mm}

${}^b${\it Kobayashi-Maskawa Institute for the Origin of Particles and the Universe, Nagoya University, \\ Furocho, Chikusa, Aichi 464-8602, Japan}

\vspace{1mm}

${}^c${\it Nishina Center for Accelerator-Based Science, RIKEN, \\ Wako 351-0198, Japan}
\end{center}

\begin{abstract}
We study QCD corrections for the associated production of a single top quark and a photon ($tq\gamma$ production) at hadron colliders. We calculate the NLO cross section at LHC and future collider energies for a variety of kinematical cuts, and we estimate uncertainties from scale dependence and from parton distributions. We also calculate differential distributions in top-quark transverse-momentum and rapidity as well as photon energy. Finally, we study higher-order corrections from soft-gluon emission for this process, and we provide approximate NNLO (aNNLO) results for the cross section and top-quark differential distributions. We also compare our calculations with recent measurements from CMS and ATLAS at the LHC and find that the aNNLO corrections improve the comparison between the data and the Standard Model predictions.
\end{abstract}

\mysection{Introduction}

The production of a single top quark in association with a photon is an interesting process for which searches have been ongoing at the LHC. Evidence for $tq\gamma$ production was first found in 13 TeV proton-proton collisions at the LHC by CMS in Ref. \cite{CMS13}. More recently, the observation of $tq\gamma$ production at 13 TeV was announced by ATLAS in Ref. \cite{ATLAS13}. The cross section for this process is sensitive to the top-quark charge and any anomalous electric and magnetic dipole moments. In addition to Standard Model (SM) couplings, anomalous $t$-$q$-$\gamma$ couplings of the top quark \cite{FK2018} with flavor-changing neutral currents may also contribute to such cross sections.

Nonstandard electric moments of fermions, i.e. anomalous magnetic moments (AMM) and electric dipole moments (EDM), are interesting observables since they may radiatively be induced by new physics beyond the SM, in contrast to the vector charges which hide these effects by renormalization. The standard way to analyze the electromagnetic moments of the top quark is to accurately measure the production of top quarks in association with a photon at colliders, and compare with the SM prediction augmented with interactions involving these anomalous moments \cite{Baur:2004uw,Fael:2013ira}. As an alternative approach, these interactions may also be probed using an off-shell photon propagating between the two incident protons \cite{Koksal:2019lyb}.

Recently, the AMM of the muon has been measured \cite{Muong-2:2021ojo}, and the combination with previous experimental data \cite{Muong-2:2006rrc} shows a deviation of more than $4\sigma$ from the SM prediction \cite{Aoyama:2020ynm}.
From a simple dimensional analysis argument, the expected tension of the AMM of the top quark due to some new physics beyond the SM with an energy scale higher than $O(100)$ GeV is \cite{Crivellin:2021spu} 
$\delta a_t \sim (m_t^2/m_\mu^2) \delta a_\mu \approx 6.7 \times 10^{-3}$,
where $\delta a_\mu $ is the deviation of the muon AMM from the SM prediction, while $m_t$ and $m_{\mu}$ are the masses of the top quark and the muon, respectively.
Of course, the scaling is strictly model dependent, and the above estimation might not work for all scenarios beyond the SM, but the value is nevertheless interesting.
From the analysis of Ref. \cite{Fael:2013ira}, the sensitivity which may be achieved by the 14 TeV run of the LHC with a luminosity of 3000 fb$^{-1}$ is $a_t \sim 0.2$.
Although reaching the above level of $\delta a_t$ with the LHC is too ambitious, the effect of top-quark AMM grows faster than the SM cross section when the center-of-mass energy is increased, so future collider experiments are definitely promising.

The chiral partner of the AMM, the EDM, is also an interesting probe of new physics beyond the SM.
As for the top quark, the interesting point is that the one-loop diagram generated by the Higgs boson with CP violating Yukawa interaction may become sizable due to the large Yukawa coupling of the top quark ($d_t \sim 10^{-17} \, \theta_H \, e$ cm, where $ \theta_H$ is the mixing angle between the CP-even and CP-odd Higgs bosons).
Currently, the top quark EDM is indirectly constrained by the experimental data of the neutron \cite{nEDM:2020crw} or the mercury atom \cite{Graner:2016ses,Sahoo:2018ile,Yanase:2020oos}, via renormalization-group analysis, as $|d_t| < 5 \cdot 10^{-20} \, e$ cm \cite{Cirigliano:2016njn}, which is tighter than the prospect of a 14 TeV LHC run with luminosity 3000 fb$^{-1}$ ($|d_t| < 10^{-17} \, e$ cm) \cite{Fael:2013ira}.
Here we note that the neutron and mercury EDMs do not only probe the top quark EDM, and they may be affected by other CP violating sources due to the compositeness of the system \cite{Yamanaka:2017mef}, so the direct measurement using colliders is absolutely necessary.
The extraction of the AMM and EDM of the top quark is only possible if we have an accurate SM prediction at hand and, thus, motivates us to directly and quantitatively study processes sensitive to them in collider experiments.

The next-to-leading-order (NLO) corrections for $tq\gamma$ production are needed for better theoretical accuracy (see also Ref. \cite{PSTZ}). In this paper, we provide a detailed study of the NLO cross section at LHC and future collider energies up to 100 TeV. We employ a variety of kinematical cuts and study the dependence of the cross section on them. We consider scale variation and uncertainties from parton distribution functions (pdf). We also calculate differential distributions in top-quark transverse momentum and rapidity as well as photon energy at LHC energies.

Soft-gluon resummation provides a powerful formalism for making theoretical predictions for perturbative corrections at higher orders \cite{NKGS1,NKGS2,KOS,NKsch,NKtW,NKtch,NKtW2016,NKNY,FK2020,FK2021}. The soft-gluon corrections appear in the perturbative series as logarithms of a variable, whose exact definition depends on the choice of kinematics, that measures the energy in the soft emission. Most calculations so far have been done for $2 \to 2$ processes, including single-top production processes (see e.g. \cite{NKsch,NKtW,NKtch,NKtW2016,NKNY}), but the resummation formalism has been recently extended in single-particle-inclusive kinematics \cite{FK2020} to processes with an arbitrary number of final-state particles. In this paper, we adapt the resummation formalism of Refs. \cite{FK2020,FK2021} for $tq\gamma$ production, and we calculate approximate NNLO (aNNLO) results for the cross section and the top-quark transverse-momentum and rapidity distributions. 

We begin in Section 2 with NLO results for $tq\gamma$ production. Numerical results are presented for LHC energies, including total cross sections and differential distributions, as well as for higher energies in future colliders. In Section 3, we discuss soft-gluon resummation and aNNLO results for this process. We conclude in Section 4.

\mysection{NLO cross sections for $tq\gamma$ production}

In this section we present results for total cross sections for $tq\gamma$ production as well as for differential distributions in top-quark transverse momentum and rapidity and in photon energy. We note that $q$ represents a light quark or antiquark. We set the top-quark mass $m_t=172.5$ GeV, and we use the latest CT18 \cite{CT18}, MSHT20 \cite{MSHT20}, and NNPDF 4.0 \cite{NNPDF4.0} pdf sets. We set the factorization and renormalization scales equal to each other and denote this common scale by $\mu$. The complete NLO results are found by using {\small \sc MadGraph5\_aMC@NLO} \cite{MG5}. 

\subsection{Total cross sections at NLO}

\begin{figure}[htbp]
\begin{center}
\includegraphics[width=86mm]{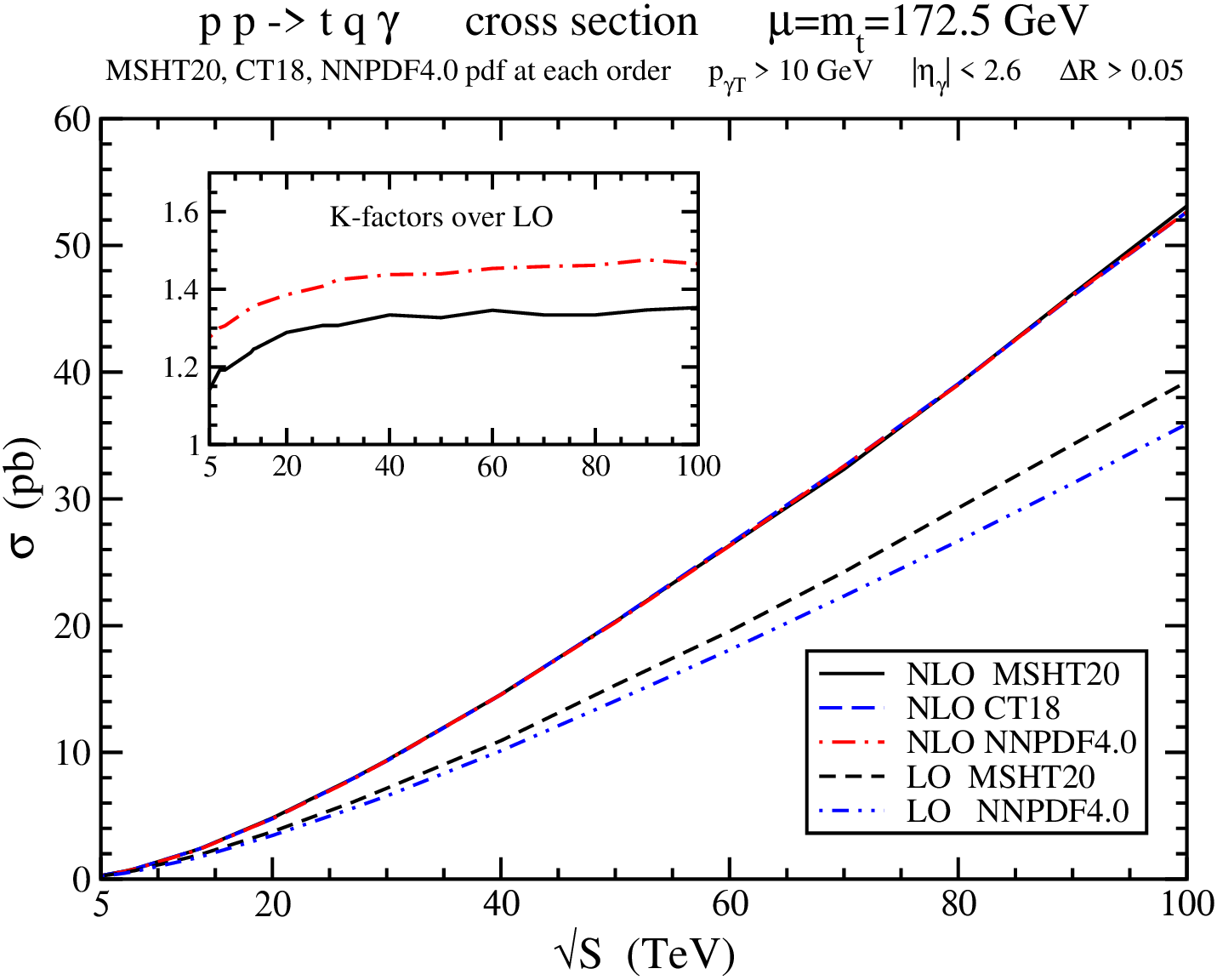}
\hspace{2mm}
\includegraphics[width=86mm]{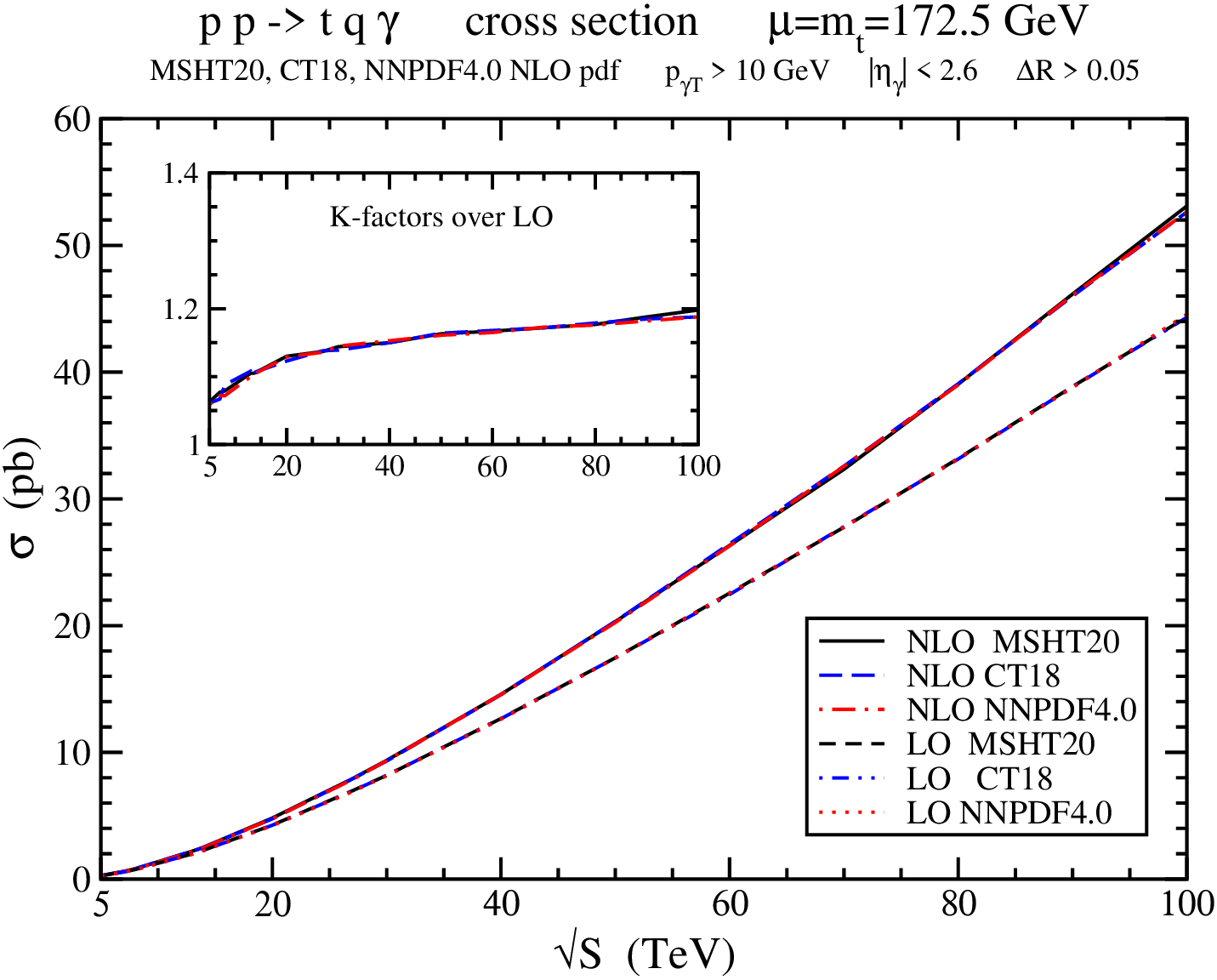}
\caption{The total cross sections at LO and NLO for $tq\gamma$ production at $pp$ colliders using various pdf sets. In the left plot we use NLO (LO) pdf for the NLO (LO) cross sections while in the right plot we use NLO pdf for all results.}
\label{tqgampdfNLO}
\end{center}
\end{figure}

In Fig. \ref{tqgampdfNLO} we display the total $pp \to tq \gamma$ cross sections with $\mu=m_t$ for collider energies up to 100 TeV. The cuts used are the following: for the photon transverse momentum, $p_{\gamma T}> 10$ GeV; for the photon pseudorapidity, $|\eta_{\gamma}| < 2.6$; and for the angular separation between the photon and the other particles, $\Delta R > 0.05$, with the definition $\Delta R=\sqrt{(\Delta \eta_{\gamma})^2+(\Delta \phi)^2}$ where $\phi$ is the azimuthal angle. These are also the cuts used for simulated samples in Ref. \cite{CMS13}. We use MSHT20 \cite{MSHT20}, CT18 \cite{CT18}, and NNPDF4.0 \cite{NNPDF4.0} pdf sets. The plot on the left shows the leading-order (LO) cross sections using LO pdf, and the NLO cross sections using NLO pdf. We see a big difference between the results with MSHT20 and NNPDF4.0 pdf at LO, but all pdf sets give similar results at NLO. The inset plot displays the $K$-factors, i.e. the ratios of the NLO cross sections to the LO ones for each pdf set. We see that the NLO corrections are important and increase with energy: the NLO/LO $K$-factor with MSHT20 pdf indicates corrections of 24\% at 13 TeV energy and 35\% at 100 TeV energy. 

The plot on the right in Fig. \ref{tqgampdfNLO} shows results for the cross section at LO and NLO using in all cases NLO pdf from the various pdf sets. This is useful in seeing how the perturbative series behaves in building the NLO result. Thus, the contribution of the LO term and the NLO corrections can be clearly identified in the final NLO result for the cross section. Both the LO and the NLO results now are nearly identical among all pdf sets. The NLO/LO (with same NLO pdf) $K$-factors are shown in the inset plot, and they increase steadily with energy, reaching 20\% at 100 TeV energy.

\begin{figure}[htbp]
\begin{center}
\includegraphics[width=86mm]{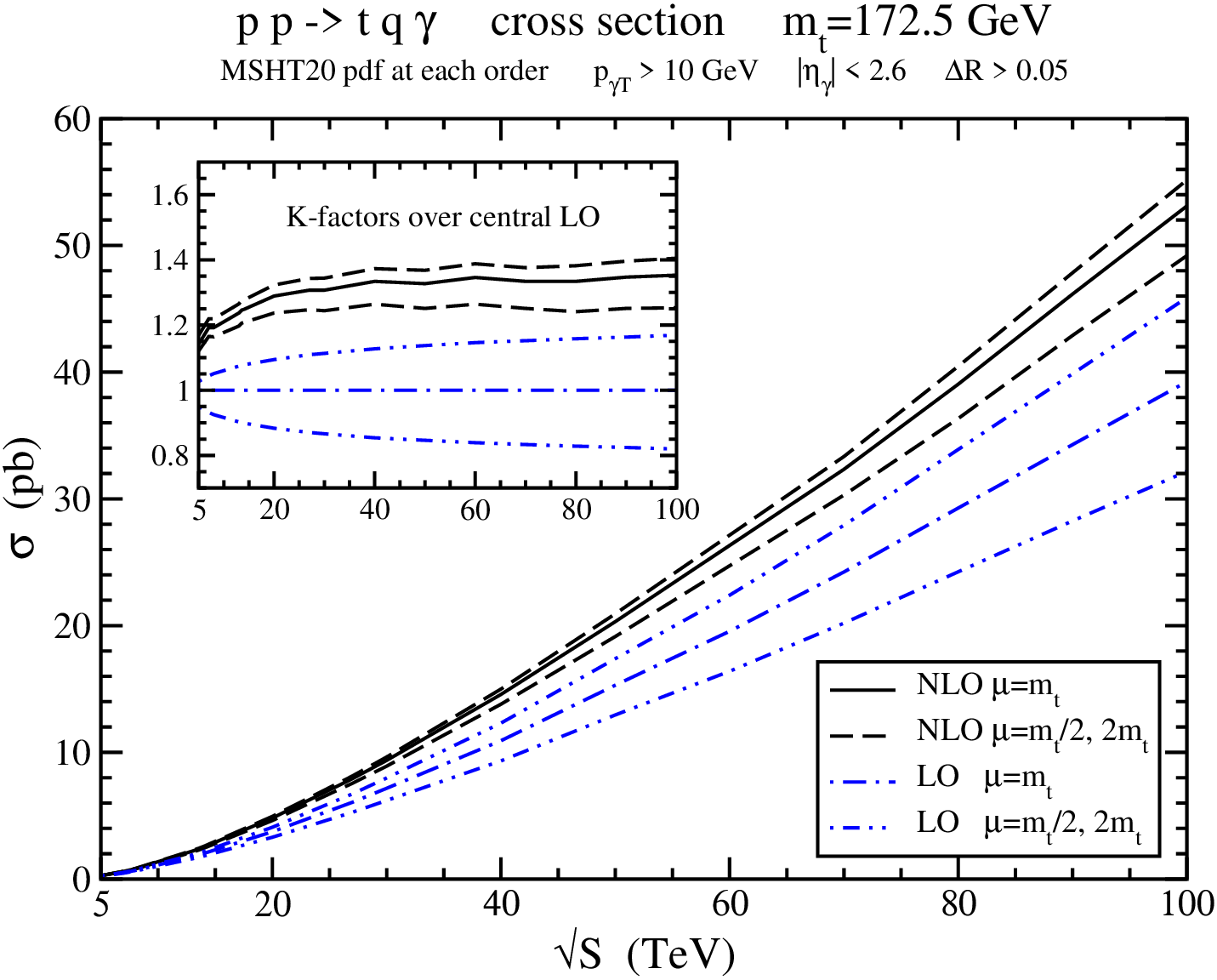}
\hspace{2mm}
\includegraphics[width=86mm]{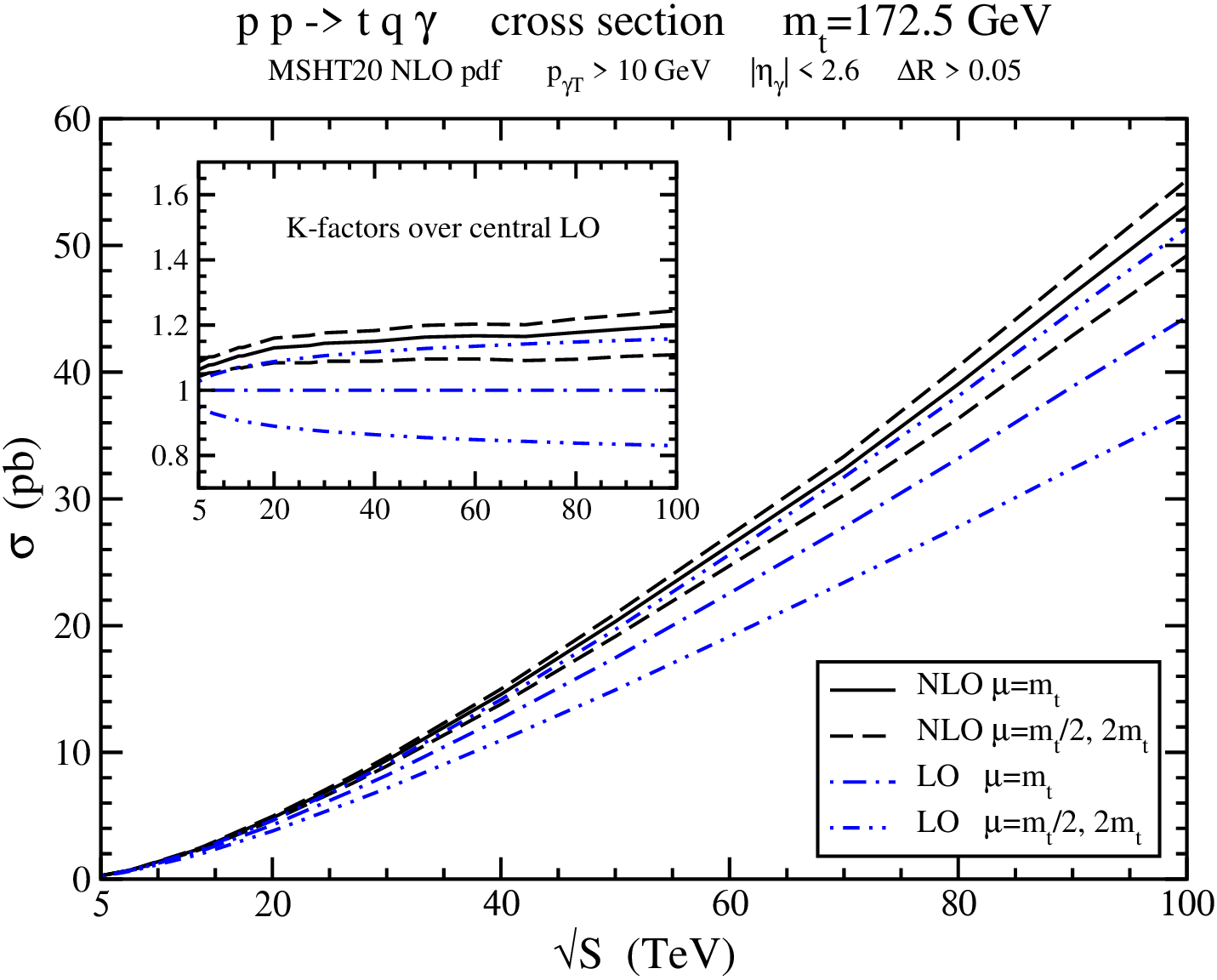}
\caption{The scale dependence of the LO and NLO total cross sections for $tq\gamma$ production at $pp$ colliders. In the left plot we use NLO (LO) pdf for the NLO (LO) cross sections while in the right plot we use NLO pdf for all results.}
\label{tqgamNLOscale}
\end{center}
\end{figure}

We continue with a study of the scale dependence of the $tq \gamma$ cross section.
In the left plot of Fig. \ref{tqgamNLOscale} we show the LO and NLO cross sections, including scale variation ($\mu=m_t/2$, $m_t$, $2m_t$) with MSHT20 pdf at each order, for $pp$-collider energies up to 100 TeV. We observe that while the LO cross section has a relatively large scale variation, this dependence is smaller at NLO. The inset plot shows ratios of the LO and NLO cross sections for various scales to the central (i.e. with $\mu=m_t$) LO cross section.

The right plot of Fig. \ref{tqgamNLOscale} shows the LO and NLO cross sections, including scale variation, with MSHT20 NLO pdf. The inset plot shows, again, ratios of the LO and NLO cross sections for various scales to the central LO cross section. Again, we observe a significantly smaller scale dependence at NLO relative to LO.

\begin{table}[htbp]
\begin{center}
\begin{tabular}{|c|c|c|c|c|c|c|c|c|} \hline
\multicolumn{9}{|c|}{$tq\gamma$ cross sections in $pp$ collisions with $p_{\gamma T}> 10$ GeV, $|\eta_{\gamma}| < 2.6$, and $\Delta R > 0.05$} \\ \hline
$\sigma$ in pb & 7 TeV & 8 TeV & 13 TeV & 13.6 TeV & 14 TeV & 27 TeV & 50 TeV & 100 TeV \\ \hline
LO (lo pdf)   & 0.508 & 0.681 & 1.78 & 1.92 & 2.02 & 6.06 & 15.3 & 39.2 \\ \hline
LO (nlo pdf)  & 0.562 & 0.753 & 1.99 & 2.16 & 2.28 & 6.97 & 17.5 & 44.3 \\ \hline
NLO (nlo pdf) & 0.605 & 0.812 & 2.20 & 2.39 & 2.52 & 7.92 & 20.3 & 53.1 \\ \hline
\end{tabular}
\caption[]{The $tq\gamma$ cross sections (in pb, with $\mu=m_t$, $p_{\gamma T}> 10$ GeV, $|\eta_{\gamma}| < 2.6$, $\Delta R > 0.05$) in $pp$ collisions with $\sqrt{S}=7$, 8, 13, 13.6, 14, 27, 50, and 100 TeV, $m_t=172.5$ GeV, and MSHT20 pdf.}
\label{table1}
\end{center}
\end{table}

In Table 1 we show total rates for $tq\gamma$ production for a variety of $pp$-collider energies at LO (using both LO and NLO MSHT20 pdf) and at NLO (using MSHT20 NLO pdf). The cuts used are the same as in the figures. The results are for LHC energies of 7, 8, 13, 13.6, and 14 TeV as well as for possible future collider energies of 27, 50, and 100 TeV. We observe an increase of two orders of magnitude in the cross section as the energy rises from 7 TeV to 100 TeV.

\begin{table}[htbp]
\begin{center}
\begin{tabular}{|c|c|c|c|c|} \hline
\multicolumn{5}{|c|}{$tq\gamma$ cross sections at NLO for LHC energies with $p_{\gamma T}> 10$ GeV, $|\eta_{\gamma}| < 2.6$, and $\Delta R > 0.05$} \\ \hline
$\sigma$ in pb & 8 TeV & 13 TeV & 13.6 TeV & 14 TeV \\ \hline
MSHT20 (nlo pdf) & $0.812^{+0.018}_{-0.019}{}^{+0.009}_{-0.008}$ & $2.20^{+0.05}_{-0.07} \pm 0.02$ & $2.39^{+0.06}_{-0.08} \pm 0.02$ & $2.52^{+0.06}_{-0.08} \pm 0.02$ \\ \hline
CT18 (nlo pdf) & $0.817^{+0.019}_{-0.021}{}^{+0.022}_{-0.019}$ & $2.20^{+0.05}_{-0.07} \pm 0.04$ & $2.37^{+0.05}_{-0.07} \pm 0.04$ & $2.52^{+0.06}_{-0.08} \pm 0.05$     \\ \hline
NNPDF4.0 (nlo pdf) & $0.794^{+0.015}_{-0.017} \pm 0.005$ & $2.16^{+0.05}_{-0.07} \pm 0.01$ & $2.35^{+0.05}_{-0.08} \pm 0.01$ & $2.49^{+0.06}_{-0.08} \pm 0.01$  \\ \hline
\end{tabular}
\caption[]{The $tq\gamma$ cross sections at NLO (in pb, with central value for $\mu=m_t$ with scale and pdf uncertainties, and using cuts $p_{\gamma T}> 10$ GeV, $|\eta_{\gamma}| < 2.6$, $\Delta R > 0.05$) in $pp$ collisions with $\sqrt{S}=8$, 13, 13.6, and 14 TeV, with $m_t=172.5$ GeV and various NLO pdf.}
\label{table2}
\end{center}
\end{table}

In Table 2 we show more detailed NLO total rates for $tq\gamma$ production for LHC energies of 8, 13, 13.6, and 14 TeV using MSHT20, CT18, and NNPDF4.0 NLO pdf sets, and the same cuts as before. We write the result with central scale $\mu=m_t$ with the first uncertainty from traditional scale variation between $m_t/2$ and $2m_t$, and the second uncertainty coming from the pdf.

The cross sections depend, of course, greatly on the choice of cuts. For the CMS cuts \cite{CMS13} of $p_{\gamma T}> 25$ GeV, $|\eta_{\gamma}| < 1.44$, $\Delta R > 0.5$ at 13 TeV and using MSHT20 NLO pdf, we find an NLO cross section of $0.553^{+0.011}_{-0.015}{}^{+0.006}_{-0.005}$ pb. However, the fractional scale uncertainties at NLO (around 2\%) and the pdf uncertainties (around 1\%) as well as the NLO/LO $K$-factor with the same NLO pdf (around 1.08) are very similar to the values with the previous cuts. 

For the ATLAS cuts \cite{ATLAS13} of $p_{\gamma T}> 20$ GeV, $|\eta_{\gamma}| < 2.37$, $\Delta R > 0.4$ at 13 TeV and using MSHT20 NLO pdf, we find an NLO cross section of $1.055^{+0.022}_{-0.029}{}^{+0.010}_{-0.009}$ pb with, again, very similar values for the fractional scale and pdf uncertainties as well as the $K$-factor.  

\begin{table}[htbp]
\begin{center}
\begin{tabular}{|c|c|c|c|c|} \hline
\multicolumn{5}{|c|}{$tq\gamma$ cross sections at 13.6 TeV LHC energy with $|\eta_{\gamma}| < 2.6$ and $\Delta R > 0.05$ } \\ \hline
$\sigma$ in pb & $p_{\gamma T}> 10$ GeV  & $p_{\gamma T}> 15$ GeV & $p_{\gamma T}> 20$ GeV & $p_{\gamma T}> 25$ GeV \\ \hline
LO (nlo pdf)  & $2.16^{+0.16}_{-0.20} \pm 0.02$ & $1.62^{+0.11}_{-0.15} \pm 0.01$ & $1.27^{+0.09}_{-0.12} \pm 0.01$ &  $1.02^{+0.07}_{-0.09} \pm 0.01$ \\ \hline
NLO (nlo pdf) & $2.39^{+0.06}_{-0.08} \pm 0.02$ & $1.81^{+0.04}_{-0.06} \pm 0.02$ & $1.40^{+0.03}_{-0.04} \pm 0.01$ & $1.14^{+0.02}_{-0.03} \pm 0.01$     \\ \hline
\end{tabular}
\caption[]{The $tq\gamma$ cross sections at LO and NLO (in pb, with central value for $\mu=m_t$ with scale and pdf uncertainties, and using cuts $p_{\gamma T}> 10$, 15, 20, and 25 GeV, as well as $|\eta_{\gamma}| < 2.6$ and $\Delta R > 0.05$) in $pp$ collisions with $\sqrt{S}=13.6$ TeV, with $m_t=172.5$ GeV and MSHT20 NLO pdf.}
\label{table3}
\end{center}
\end{table}

Next, we investigate further the dependence of the cross section on kinematical cuts.
In Table 3, we present the LO and NLO cross sections (all with MSHT20 NLO pdf) at 13.6 TeV energy, with $|\eta_{\gamma}| < 2.6$ and $\Delta R > 0.05$, for four different  $p_{\gamma T}$ cuts. The scale and pdf uncertainties are also provided. Although the values of the cross sections diminish quickly with increasing cut on $p_T$, the $K$-factors remain almost identical. The fractional scale and pdf uncertainties also remain quite similar.

\begin{table}[htbp]
\begin{center}
\begin{tabular}{|c|c|c|c|c|} \hline
\multicolumn{5}{|c|}{$tq\gamma$ cross sections at 13.6 TeV LHC energy with $p_{\gamma T} > 10$ GeV and $\Delta R > 0.05$ } \\ \hline
$\sigma$ in pb & $|\eta_{\gamma}| < 1.5 $  & $|\eta_{\gamma}| < 2.0$ & $|\eta_{\gamma}| <  2.6$ & $|\eta_{\gamma}| < 3.0$ \\ \hline
LO (nlo pdf)  & $1.31^{+0.10}_{-0.13} \pm 0.01$ & $1.72^{+0.13}_{-0.16} \pm 0.01$ & $2.16^{+0.16}_{-0.20} \pm 0.02$ & $2.40^{+0.17}_{-0.23} \pm 0.02 $ \\ \hline
NLO (nlo pdf) & $1.48^{+0.04}_{-0.05} \pm 0.01 $ & $1.91^{+0.04}_{-0.06} \pm 0.02$ & $2.39^{+0.06}_{-0.08} \pm 0.02$ &  $2.64^{+0.06}_{-0.08} \pm 0.02 $  \\ \hline
\end{tabular}
\caption[]{The $tq\gamma$ cross sections at LO and NLO (in pb, with central value for $\mu=m_t$ with scale and pdf uncertainties, and using cuts $|\eta_{\gamma}| < 1.5$, 2.0, 2.6, and 3.0, as well as $p_{\gamma T} > 10$ GeV and $\Delta R > 0.05$) in $pp$ collisions with $\sqrt{S}=13.6$ TeV, with $m_t=172.5$ GeV and MSHT20 NLO pdf.}
\label{table4}
\end{center}
\end{table}

In Table 4, we present the LO and NLO cross sections (all with MSHT20 NLO pdf) at 13.6 TeV energy, with $p_{\gamma T} > 10$ GeV and $\Delta R > 0.05$, for four different $|\eta_{\gamma}|$ cuts. We also show the scale and pdf uncertainties in all results. The values of the cross sections increase quickly with increasing upper bound on $|\eta_{\gamma}|$, but the $K$-factors and the fractional scale and pdf uncertainties remain quite similar.

\begin{table}[htbp]
\begin{center}
\begin{tabular}{|c|c|c|c|c|} \hline
\multicolumn{5}{|c|}{$tq\gamma$ cross sections at 13.6 TeV LHC energy with $p_{\gamma T} > 10$ GeV and $|\eta_{\gamma}| < 2.6$ } \\ \hline
$\sigma$ in pb & $\Delta R > 0.05$ & $\Delta R > 0.1$ & $\Delta R > 0.3$ & $\Delta R > 0.5$  \\ \hline
LO (nlo pdf) & $2.16^{+0.16}_{-0.20} \pm 0.02$ & $2.10^{+0.15}_{-0.20} \pm 0.02 $ & $1.99^{+0.14}_{-0.19} \pm 0.02 $ & $1.93^{+0.13}_{-0.18} \pm 0.02 $ \\ \hline
NLO (nlo pdf) & $2.39^{+0.06}_{-0.08} \pm 0.02$ & $2.31^{+0.05}_{-0.07} \pm 0.02 $ & $2.12^{+0.04}_{-0.06} \pm 0.02 $ &  $ 2.02^{+0.04}_{-0.06} \pm 0.02 $  \\ \hline
\end{tabular}
\caption[]{The $tq\gamma$ cross sections at LO and NLO (in pb, with central value for $\mu=m_t$ with scale and pdf uncertainties, and using cuts $\Delta R > 0.05$, 0.1, 0.3, and 0.5, as well as $p_{\gamma T} > 10$ GeV and $|\eta_{\gamma}| < 2.6$) in $pp$ collisions with $\sqrt{S}=13.6$ TeV, with $m_t=172.5$ GeV and MSHT20 NLO pdf.}
\label{table5}
\end{center}
\end{table}

In Table 5, we present the LO and NLO cross sections (all with MSHT20 NLO pdf) at 13.6 TeV energy, with $p_{\gamma T} > 10$ GeV and $|\eta_{\gamma}| < 2.6$, for four different $\Delta R$ cuts. Again, we show the scale and pdf uncertainties in all results. The values of the cross sections decrease with increasing cut on $\Delta R$. The $K$-factor for the largest cut is significantly smaller than at the lower values for the cut. The fractional scale and pdf uncertainties remain quite similar throughout.

\begin{table}[htbp]
\begin{center}
\begin{tabular}{|c|c|c|c|c|c|c|c|c|} \hline
\multicolumn{9}{|c|}{${\bar t} q\gamma$ cross sections in $pp$ collisions with $p_{\gamma T}> 10$ GeV, $|\eta_{\gamma}| < 2.6$, and $\Delta R > 0.05$} \\ \hline
$\sigma$ in pb & 7 TeV & 8 TeV & 13 TeV & 13.6 TeV & 14 TeV & 27 TeV & 50 TeV & 100 TeV \\ \hline
LO (lo pdf)   & 0.361 & 0.490 & 1.35 & 1.47 & 1.56 & 5.05 & 13.4 & 35.9 \\ \hline
LO (nlo pdf)  & 0.348 & 0.478 & 1.36 & 1.49 & 1.58 & 5.26 & 14.1 & 38.3 \\ \hline
NLO (nlo pdf) & 0.383 & 0.532 & 1.53 & 1.69 & 1.80 & 6.16 & 16.7 & 45.7 \\ \hline
\end{tabular}
\caption[]{The ${\bar t} q\gamma$ cross sections (in pb, with $\mu=m_t$, $p_{\gamma T}> 10$ GeV, $|\eta_{\gamma}| < 2.6$, $\Delta R > 0.05$) in $pp$ collisions with $\sqrt{S}=7$, 8, 13, 13.6, 14, 27, 50, and 100 TeV, $m_t=172.5$ GeV, and MSHT20 pdf.}
\label{table6}
\end{center}
\end{table}

Finally, we briefly discuss the case of antitop associated production with a photon, ${\bar t} q \gamma$. We note that the cross sections for the case of an antitop are different (somewhat smaller) due to the different pdf involved in the production processes.

In Table 6 we show total rates for ${\bar t}q\gamma$ production for a variety of $pp$-collider energies at LO (using both LO and NLO MSHT20 pdf) and at NLO (using MSHT20 NLO pdf). The cuts used are the same as in Table 1. Results are given for LHC energies of 7, 8, 13, 13.6, and 14 TeV as well as for possible future collider energies of 27, 50, and 100 TeV. We observe an increase of two orders of magnitude in the ${\bar t} q \gamma$ cross section as the energy rises from 7 TeV to 100 TeV. The NLO ${\bar t}q\gamma$ cross section with scale and pdf uncertainties using MSHT20 NLO pdf is $1.53^{+0.04}_{-0.05} \pm 0.02$ pb at 13 TeV; $1.69^{+0.05}_{-0.06} \pm 0.02$ pb at 13.6 TeV; and $1.80^{+0.05}_{-0.07} \pm 0.02$ pb at 14 TeV.

For the CMS cuts \cite{CMS13} of $p_{\gamma T}> 25$ GeV, $|\eta_{\gamma}| < 1.44$, $\Delta R > 0.5$ at 13 TeV and using MSHT20 NLO pdf, we find a ${\bar t}q\gamma$ production cross section at NLO of $0.383^{+0.009}_{-0.012}{}^{+0.005}_{-0.004}$ pb. 

For the ATLAS cuts \cite{ATLAS13} of $p_{\gamma T}> 20$ GeV, $|\eta_{\gamma}| < 2.37$, $\Delta R > 0.4$ at 13 TeV and using MSHT20 NLO pdf, we find a ${\bar t}q\gamma$ production cross section at NLO of $0.711^{+0.019}_{-0.024}{}^{+0.008}_{-0.007}$ pb.

\subsection{Top-quark $p_T$ and rapidity distributions at NLO}

Next, we present top-quark transverse-momentum ($p_{tT}$) and rapidity ($y_t$) distributions in $tq\gamma$ production, which provide more information than total cross sections. 

\begin{figure}[htbp]
\begin{center}
\includegraphics[width=58mm]{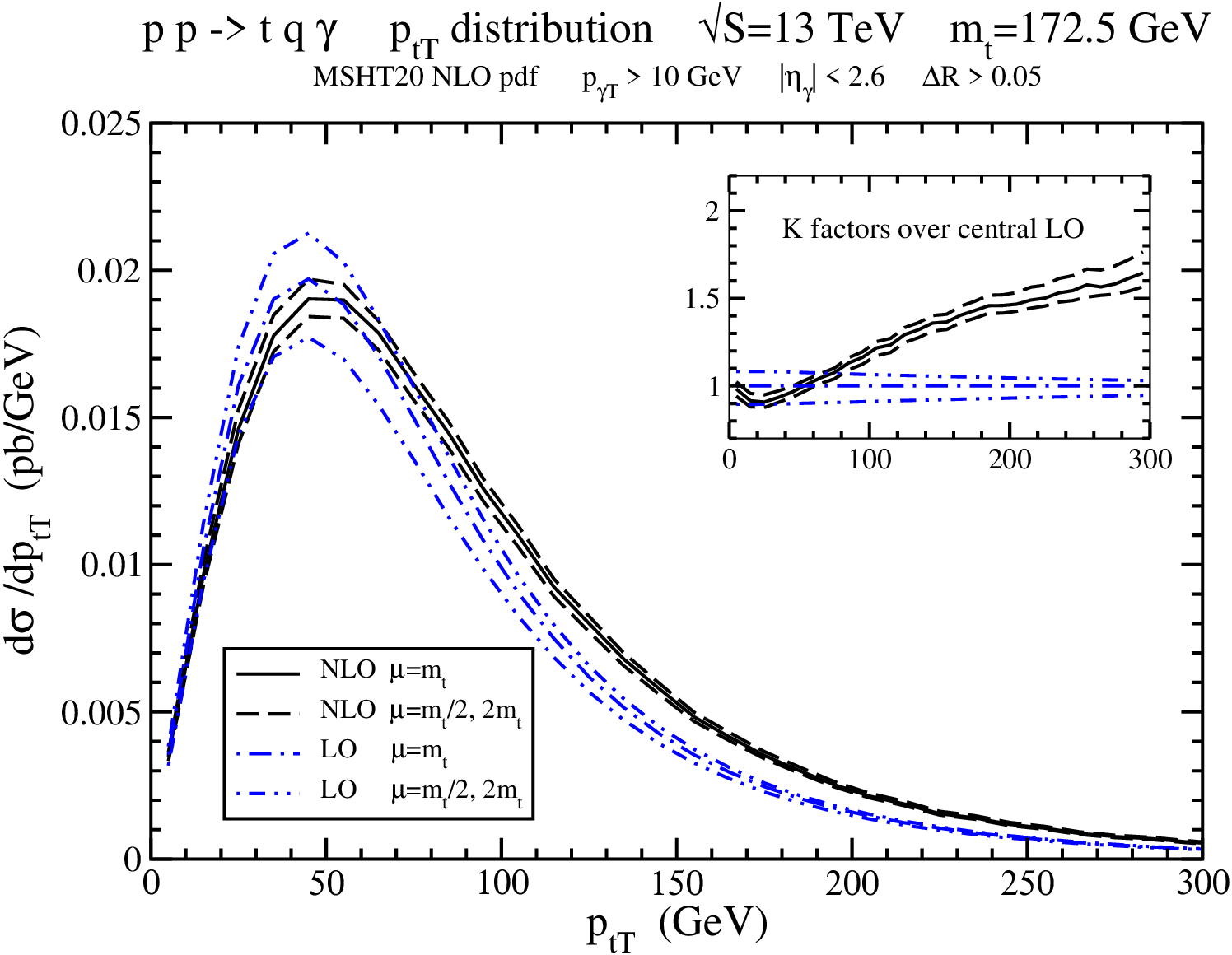}
\includegraphics[width=58mm]{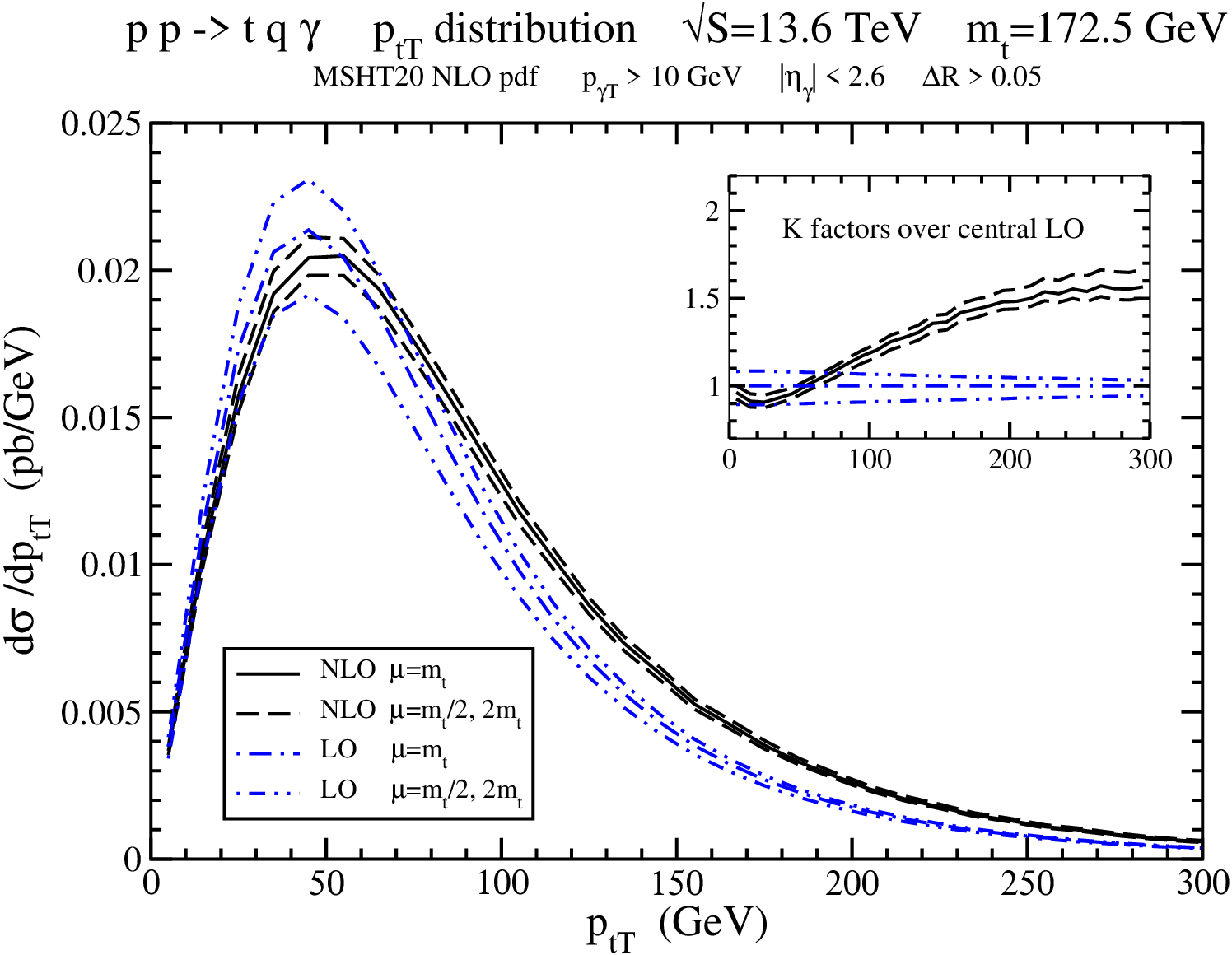}
\includegraphics[width=58mm]{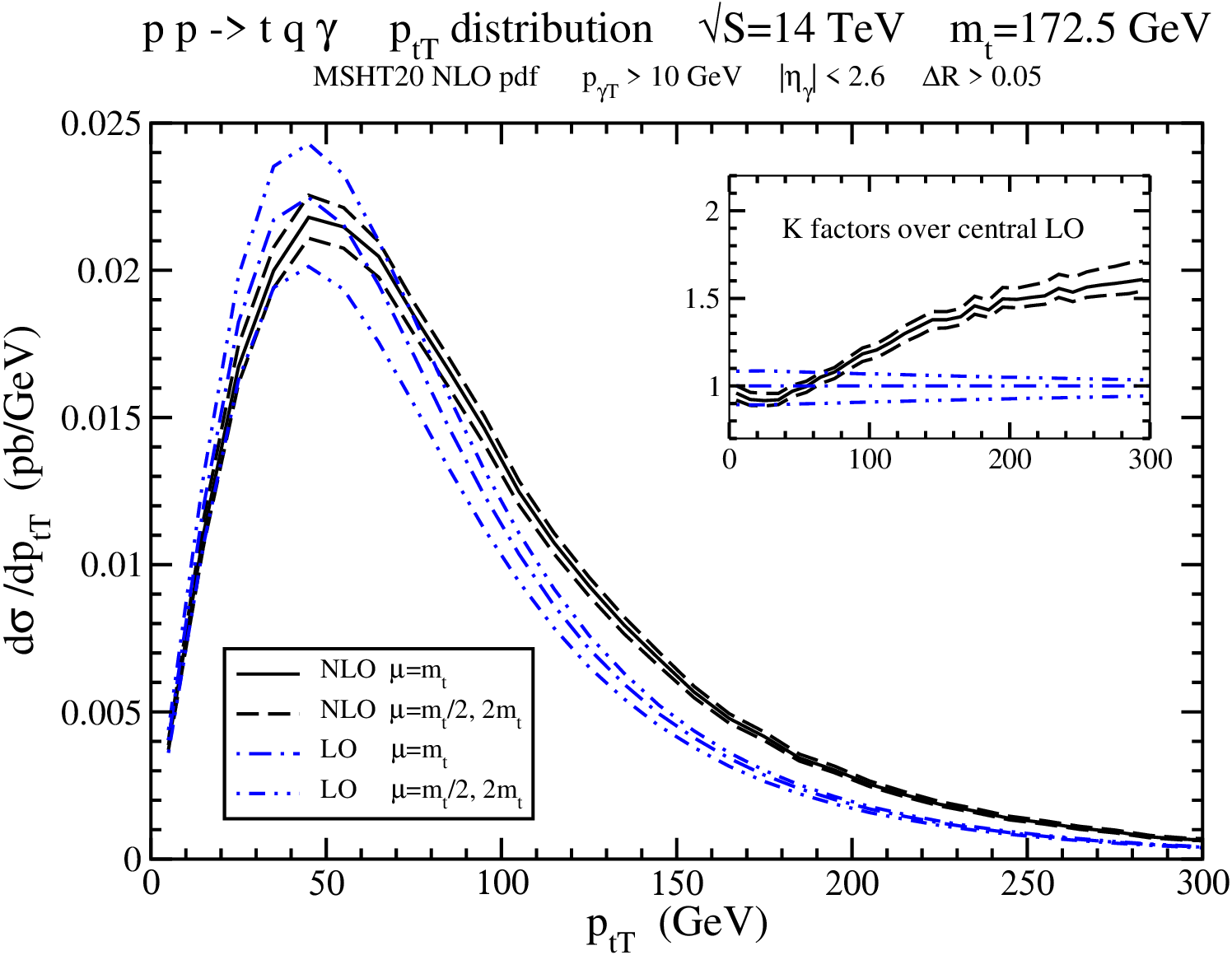}
\caption{The LO and NLO top-quark $p_T$ distributions in $tq\gamma$ production at (left) 13 TeV, (middle) 13.6 TeV, and (right) 14 TeV LHC energies.}
\label{pTtqgamma}
\end{center}
\end{figure}

In Fig. \ref{pTtqgamma} we present the top-quark $p_{tT}$ distributions, $d\sigma/dp_{tT}$, at 13 TeV (left), 13.6 TeV (middle), and 14 TeV (right) energies using MSHT20 NLO pdf. We show LO and NLO distributions for three different choices of scale. We observe that the NLO distributions peak at a $p_{tT}$ value of around 50 GeV and quickly diminish at high $p_{tT}$ values. The inset plots show the $K$ factors with respect to the central ($\mu=m_t$) LO $p_{tT}$ distribution. We observe that the $K$ factors increase with top-quark $p_{tT}$, reaching the value of 1.6 for the central NLO/LO result at a $p_{tT}$ of 300 GeV. We also observe a reduced scale dependence at NLO particularly at small and medium $p_{tT}$ values.

\begin{figure}[htbp]
\begin{center}
\includegraphics[width=57mm]{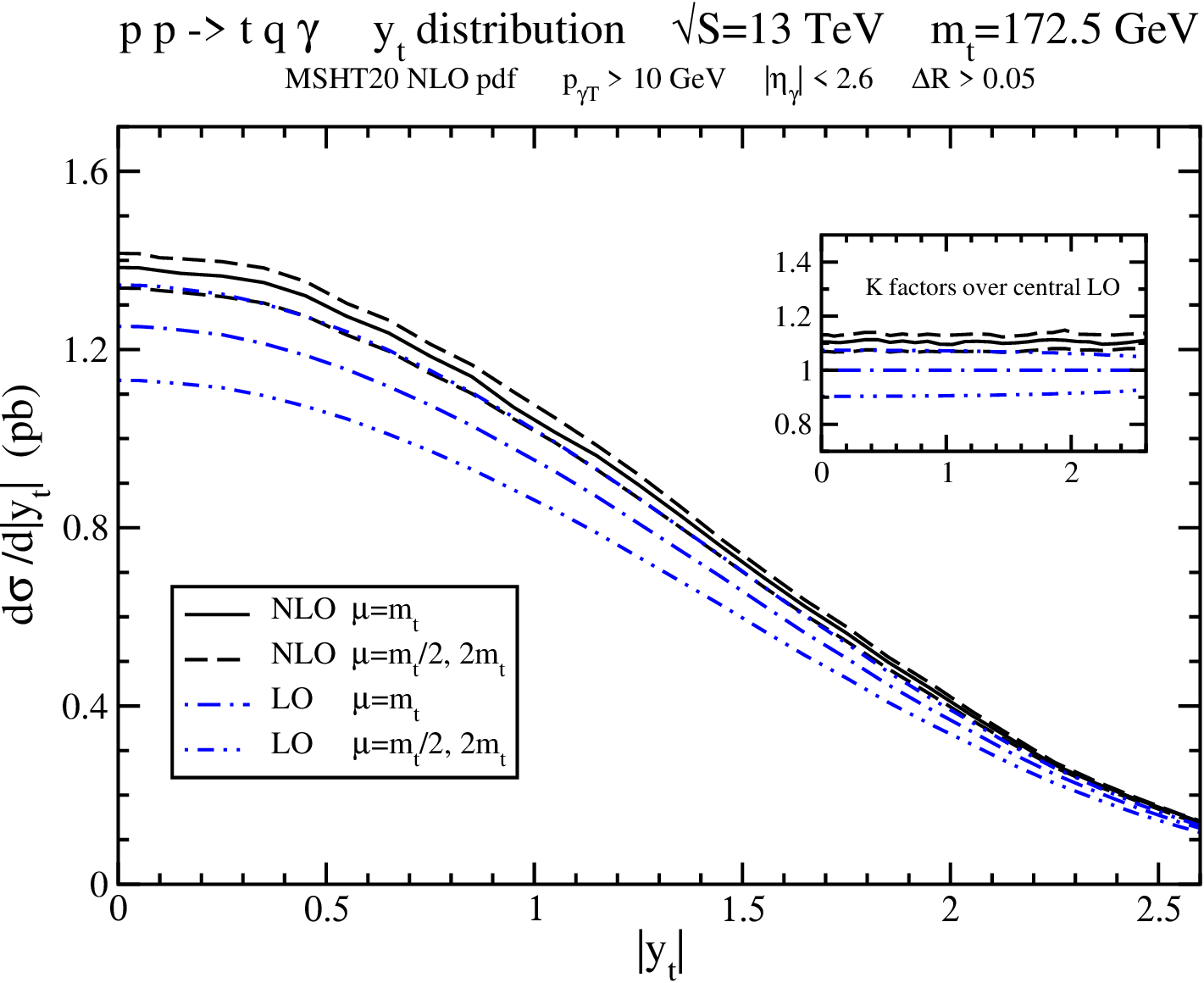}
\hspace{1mm}
\includegraphics[width=57mm]{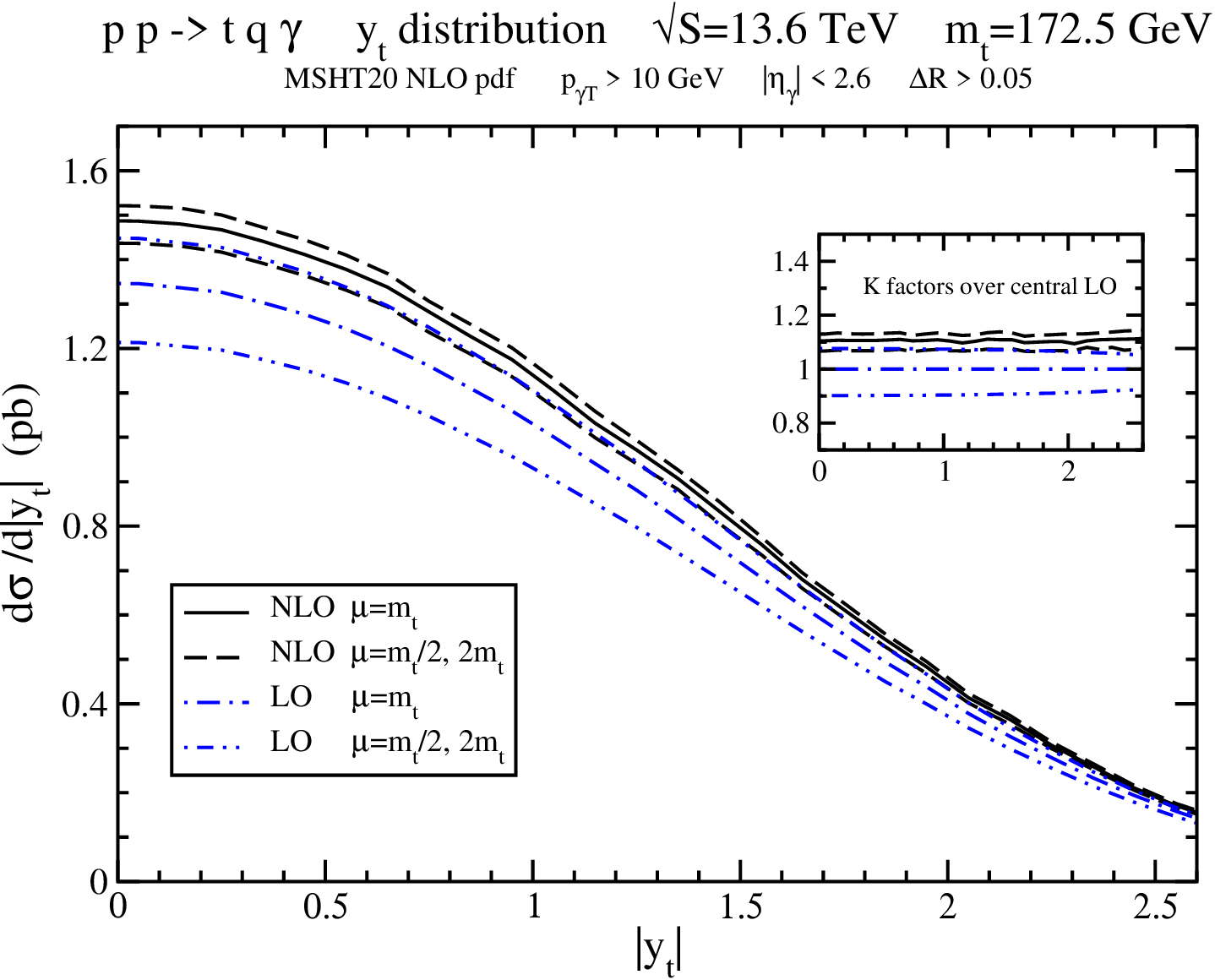}
\hspace{1mm}
\includegraphics[width=57mm]{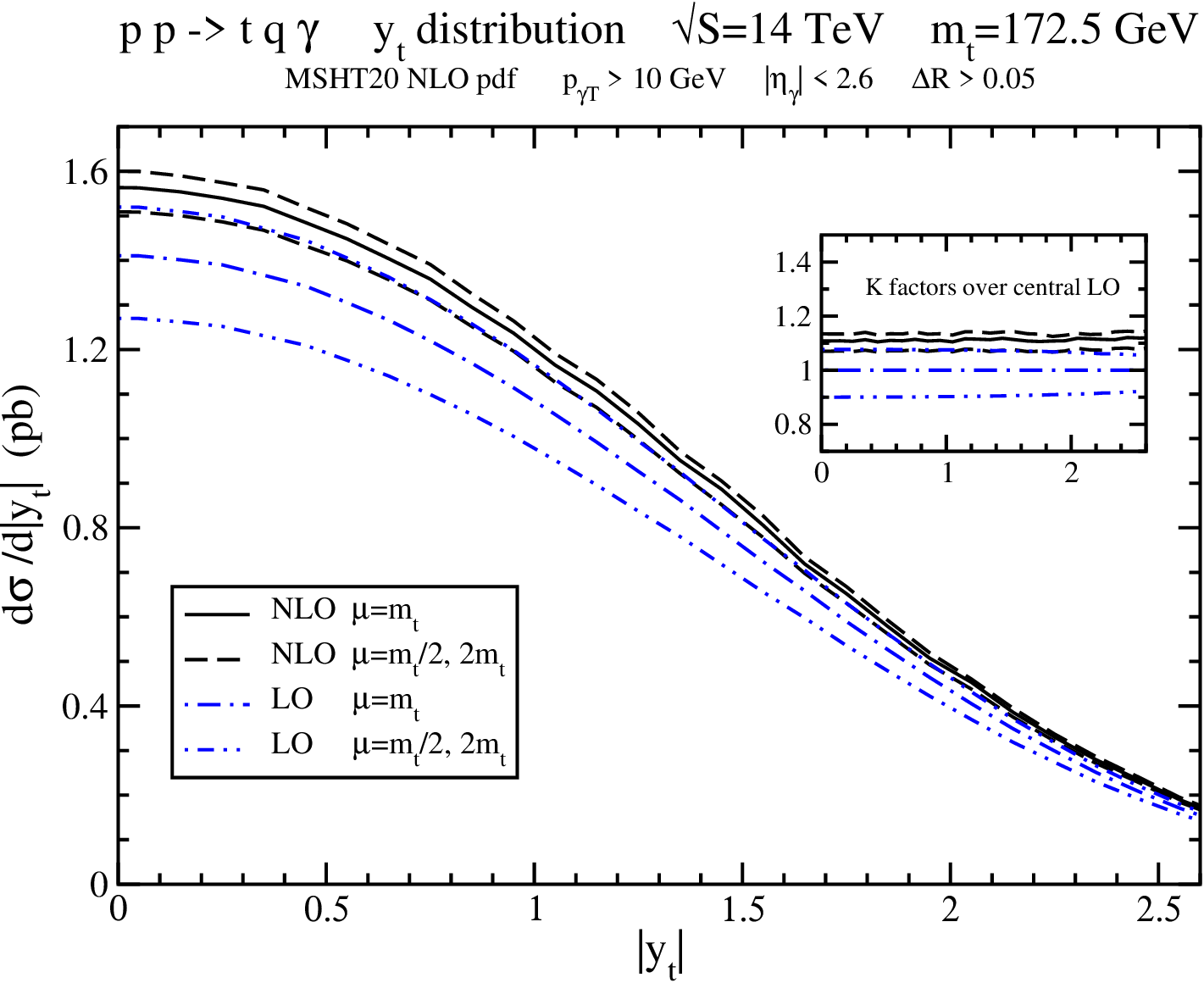}
\caption{The LO and NLO top-quark rapidity distributions in $tq\gamma$ production at (left) 13 TeV, (middle) 13.6 TeV, and (right) 14 TeV LHC energies.}
\label{ytqgamma}
\end{center}
\end{figure}

In Fig. \ref{ytqgamma} we present the top-quark rapidity distributions, $d\sigma/d|y_t|$, at 13 TeV (left), 13.6 TeV (middle), and 14 TeV (right) energies using MSHT20 NLO pdf. Again, we show LO and NLO distributions for three different choices of scale. The inset plots indicate that the $K$ factors with respect to the central ($\mu=m_t$) LO rapidity distribution show little variation over the rapidity range. The central NLO/LO $K$-factor is around 1.1. We also find a much reduced scale dependence at NLO relative to LO throughout the rapidity range shown.

\subsection{Photon energy distributions at NLO}

\begin{figure}[htbp]
\begin{center}
\includegraphics[width=57mm]{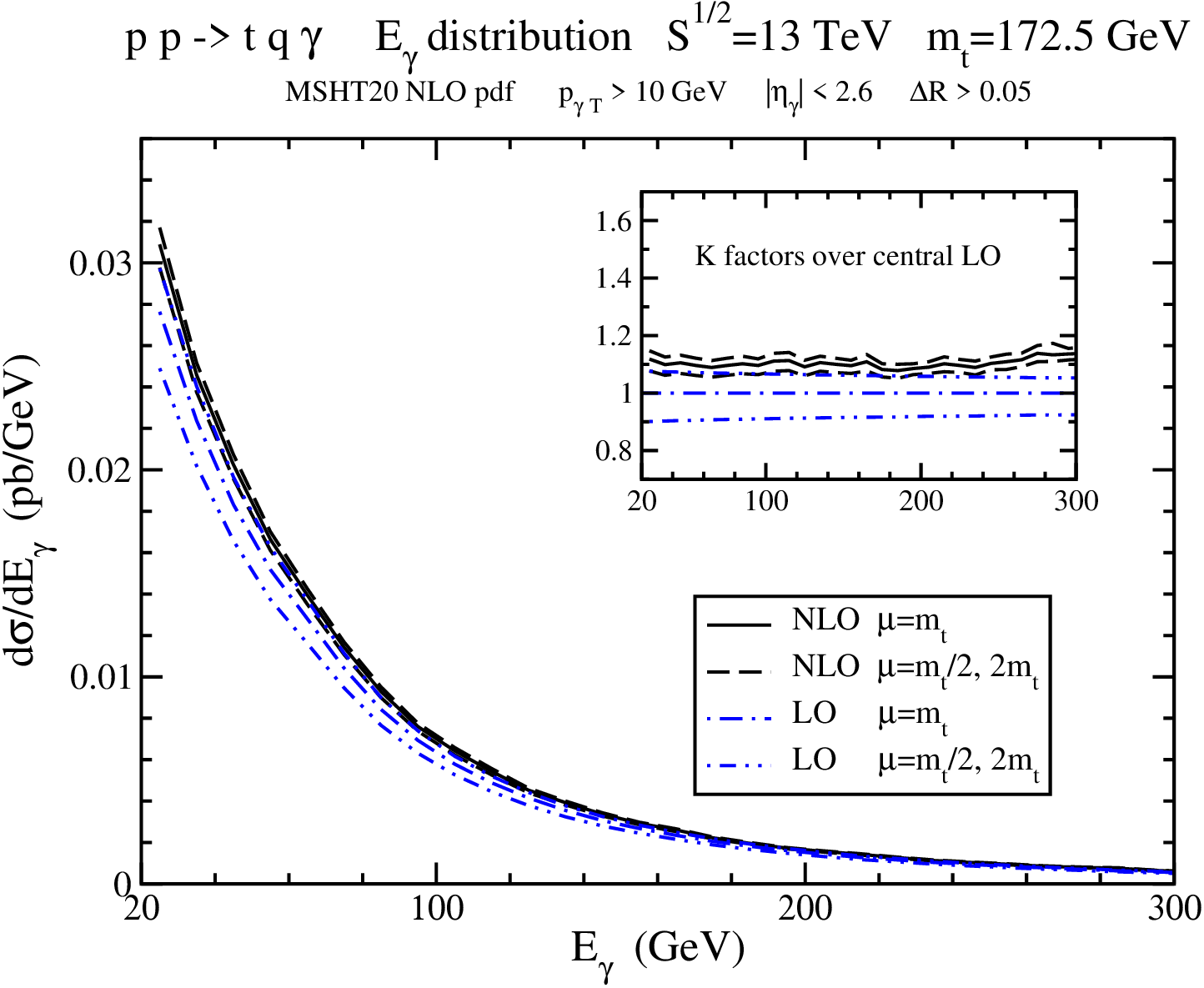}
\hspace{1mm}
\includegraphics[width=57mm]{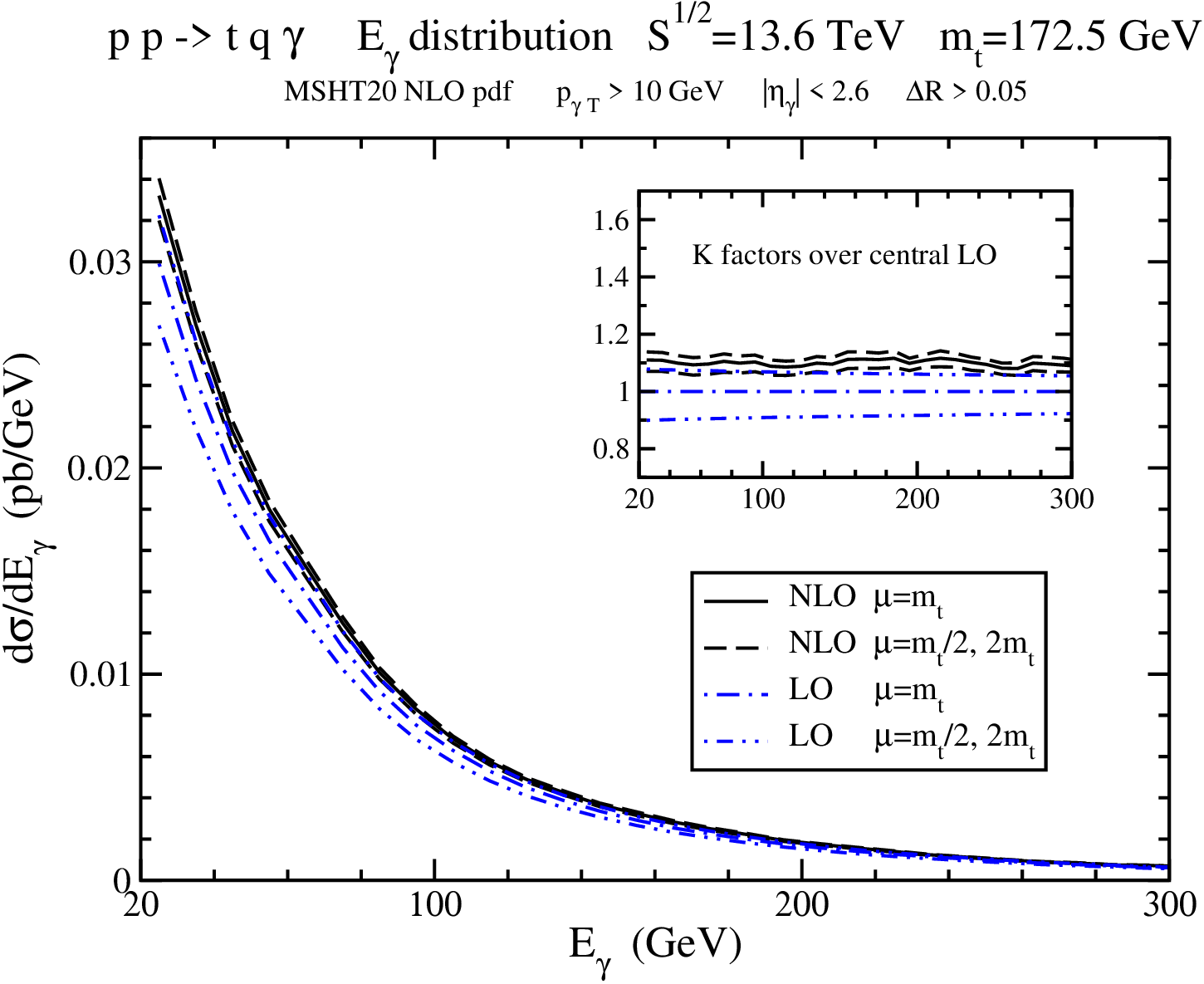}
\hspace{1mm}
\includegraphics[width=57mm]{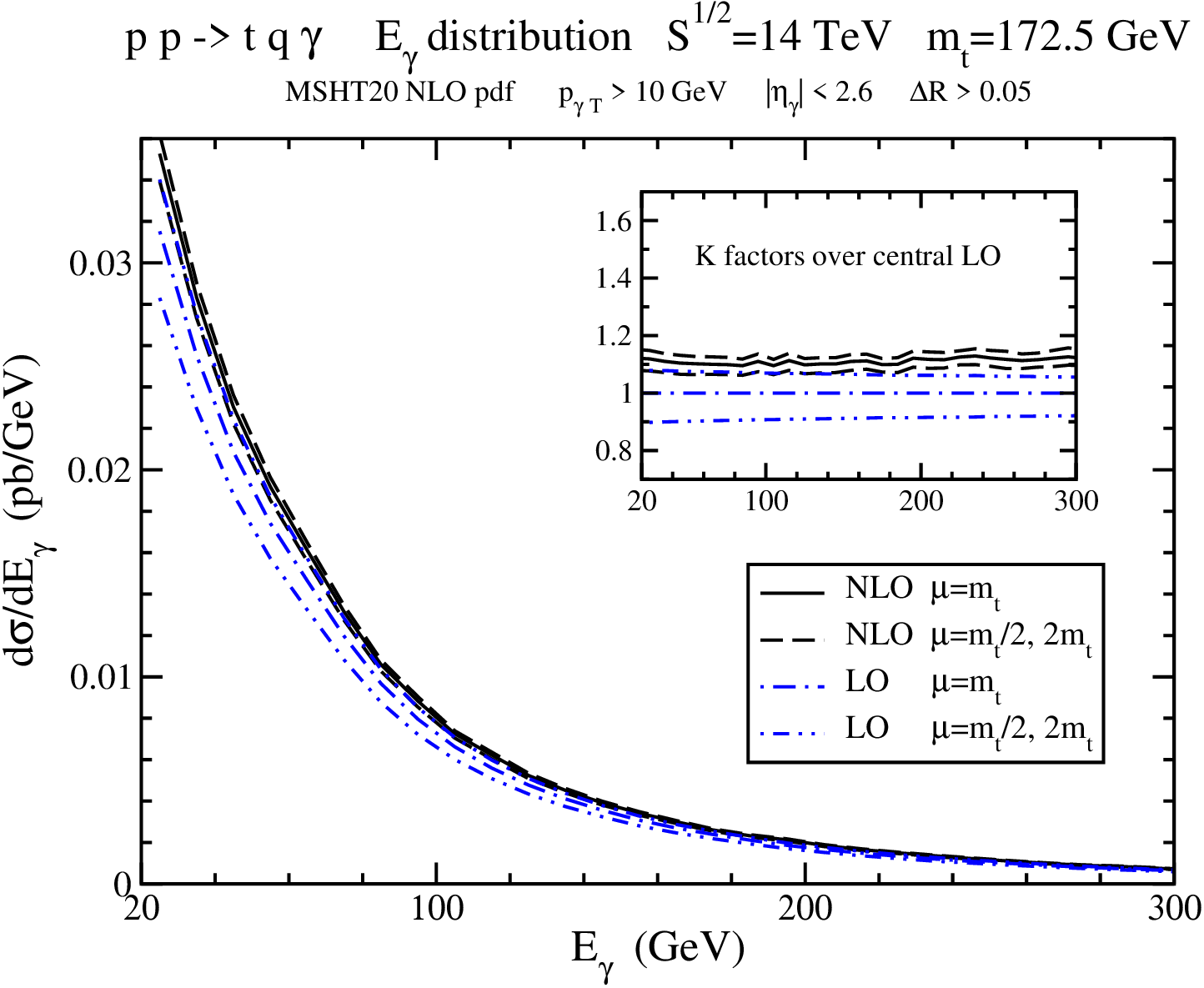}
\caption{The LO and NLO photon energy distributions in $tq\gamma$ production at (left) 13 TeV, (middle) 13.6 TeV, and (right) 14 TeV LHC energies.}
\label{Etqgamma}
\end{center}
\end{figure}

We continue with the energy distribution of the photon, $d\sigma/dE_{\gamma}$, in $tq\gamma$ production. In Fig. \ref{Etqgamma} we present the photon energy distributions at 13 TeV (left), 13.6 TeV (middle), and 14 TeV (right) energies using MSHT20 NLO pdf. The distributions fall drastically with photon energy in all cases. As before, we show LO and NLO distributions for three different choices of scale. The scale dependence is evidently much reduced at NLO relative to LO.

The inset plots show the $K$ factors with respect to the central ($\mu=m_t$) LO energy distribution. The NLO corrections provide significant enhancements. The central NLO/LO $K$-factor remains relatively stable in the energy range shown, with a value around 1.1.

\mysection{Soft-gluon resummation for $tq\gamma$ production}

In this section we discuss the formalism for soft-gluon resummation for $tq\gamma$ production, adapting the results from \cite{FK2020,FK2021}. Soft-gluon resummation \cite{NKGS1,NKGS2,KOS,NKsch,NKtW,NKtch,NKtW2016,NKNY,FK2020,FK2021} has long been known to be very important for hard-scattering processes, particularly for top-quark processes, since the cross section receives large soft-gluon corrections near partonic threshold due to the large mass of the top quark. This has been known for many top-quark processes, including top-antitop pair production, single-top production in the $s$-, $t$-, and $tW$ channels; $tqZ$ production; $tqH$ production; and even processes beyond the SM, including $tH^-$ production and processes involving top-quark anomalous couplings, such as $t\gamma$, $tZ$, $tZ'$, and $tg$ production (see e.g. the review in Ref. \cite{NKtoprev}). In all these processes, including the process $t q \gamma$ of this study, the soft-gluon corrections are dominant and account for the majority of the complete corrections at NLO. 

Furthermore, for those processes for which complete NNLO corrections are known, namely top-antitop production and $s$-channel single-top production, it is known that the soft-gluon corrections are also dominant at NNLO. In fact, the soft-gluon calculations at NNLO, with explicit analytical and numerical results, predated and numerically predicted very well the later complete NNLO results; this has been discussed in particular for top-antitop production in many papers, see e.g. the review in Ref. \cite{NKtoprev}. This provides strong motivation for the study of resummation for $tq\gamma$ production, which has not been done before.

\subsection{Resummation formalism}

We consider the partonic process $a(p_a)+b(p_b) \to t(p_t)+q(p_q)+\gamma(p_{\gamma})$. We define the parton-level kinematical variables $s=(p_a+p_b)^2$, $t=(p_a-p_t)^2$, and $u=(p_b-p_t)^2$. If an additional gluon is emitted with momentum $p_g$ in the final state, then momentum conservation gives $p_a +p_b=p_t +p_q +p_{\gamma}+p_g$. We then define a threshold variable $s_4=(p_q+p_{\gamma}+p_g)^2-(p_q+p_{\gamma})^2$ which describes the extra energy from gluon emission and vanishes as $p_g \to 0$. An equivalent definition is $s_4=s+t+u-p_t^2-(p_q+p_{\gamma})^2$.

We write the differential cross section for $tq\gamma$ production in proton-proton collisions as a convolution, 
\beq
E_t\frac{d\sigma_{pp \to tq\gamma}}{d^3p_t}=\sum_{a,b} \; 
\int dx_a \, dx_b \,  \phi_{a/p}(x_a, \mu_F) \, \phi_{b/p}(x_b, \mu_F) \, 
E_t \frac{d{\hat \sigma}_{ab \to tq\gamma}(s_4, \mu_F)}{d^3p_t} \, ,
\label{sigma}
\eeq
where $E_t$ is the energy of the observed top quark, $\mu_F$ is the factorization scale, $\phi_{a/p}$  and $\phi_{b/p}$ are parton distribution functions for parton $a$ and parton $b$, respectively, in the proton, and ${\hat \sigma}_{ab \to tq\gamma}$ is the hard-scattering partonic cross section.

The cross section factorizes under transforms \cite{FK2020}. We define Laplace transforms of the partonic cross section as ${\tilde{\hat\sigma}}_{ab \to tq\gamma}(N)=\int_0^s (ds_4/s) \,  e^{-N s_4/s} \, {\hat\sigma}_{ab \to tq\gamma}(s_4)$, where $N$ is the transform variable. The logarithms of $s_4$ in the perturbative series transform into logarithms of $N$, which exponentiate. We also define transforms of the pdf via ${\tilde \phi}(N)=\int_0^1 e^{-N(1-x)} \phi(x) \, dx$. Replacing the colliding protons by partons in Eq. (\ref{sigma}) \cite{NKGS2,GS}, we thus have under transforms the factorized form
\beq
E_t\frac{d{\tilde \sigma}_{ab \to tq\gamma}(N)}{d^3p_t}= {\tilde \phi}_{a/a}(N_a, \mu_F) \, {\tilde \phi}_{b/b}(N_b, \mu_F) \, E_t \frac{d{\tilde{\hat \sigma}}_{ab \to tq\gamma}(N, \mu_F)}{d^3p_t} \, .
\label{factorized}
\eeq

The cross section can be refactorized \cite{NKGS1,NKGS2,KOS,FK2020,FK2021} in terms of a hard function, $H_{ab \to tq\gamma}$, which is infrared safe, and a soft function, $S_{ab \to tq\gamma}$, which describes the emission of noncollinear soft gluons. Both the hard and the soft functions are $2\times 2$ matrices in the color space of the partonic scattering. We have
\beq
E_t\frac{d{\tilde{\sigma}}_{ab \to tq\gamma}(N)}{d^3p_t}={\tilde \psi}_{a/a}(N_a,\mu_F) \, {\tilde \psi}_{b/b}(N_b,\mu_F) \, {\tilde J}_q (N, \mu_F) \, {\rm tr} \left\{H_{ab \to tq\gamma} \left(\alpha_s(\mu_R)\right) \, {\tilde S}_{ab \to tq\gamma} \left(\frac{\sqrt{s}}{N \mu_F} \right)\right\} \, ,
\label{refactorized}
\eeq
where the functions $\psi$ are distributions for incoming partons at fixed value of momentum and involve collinear emission \cite{NKGS1,NKGS2,KOS,GS} while the function $J_q$ describes radiation from the final-state light quark.

Comparing Eqs. (\ref{factorized}) and (\ref{refactorized}), we find an expression for the hard-scattering partonic cross section in transform space
\beq
E_t\frac{d{\tilde{\hat \sigma}}_{ab \to tq\gamma}(N)}{d^3p_t}=
\frac{{\tilde \psi}_{a/a}(N_a, \mu_F) \, {\tilde \psi}_{b/b}(N_b, \mu_F) \, {\tilde J_q} (N, \mu_F)}{{\tilde \phi}_{a/a}(N_a, \mu_F) \, {\tilde \phi}_{b/b}(N_b, \mu_F)} \; \,  {\rm tr} \left\{H_{ab \to tq\gamma}\left(\alpha_s(\mu_R)\right) \, 
{\tilde S}_{ab \to tq\gamma}\left(\frac{\sqrt{s}}{N \mu_F} \right)\right\} \, .
\label{sigN}
\eeq

The dependence of the soft matrix on the transform variable, $N$, is resummed via renormalization-group evolution \cite{NKGS1,NKGS2}. Thus, ${\tilde S}_{ab \to tq\gamma}$ obeys a renormalization-group equation in terms of a soft anomalous dimension matrix, $\Gamma_{\! S \, ab \to tq\gamma}$, which is calculated from the coefficients of the ultraviolet poles of the relevant eikonal diagrams \cite{NKGS1,NKGS2,KOS,FK2020,NKsch,NKtW,NKtch,NK2loop,NKtt2l,NK3loop}.

The $N$-space resummed cross section is derived from the renormalization-group evolution of the $N$-dependent functions in Eq. (\ref{sigN}), i.e. ${\tilde S}_{ab \to tq\gamma}$, ${\tilde \psi}$, ${\tilde \phi}$, and ${\tilde J}_q$, and it is given by
\beqa
E_t\frac{d{\tilde{\hat \sigma}}_{ab \to tq\gamma}^{\rm resum}(N)}{d^3p_t} &=&
\exp\left[\sum_{i=a,b} E_{i}(N_i)\right] \, 
\exp\left[\sum_{i=a,b} 2 \int_{\mu_F}^{\sqrt{s}} \frac{d\mu}{\mu} \gamma_{i/i}(N_i)\right] \, 
\exp\left[E'_q(N)\right]
\nonumber\\ && \hspace{-5mm} \times \,
{\rm tr} \left\{H_{ab \to tq\gamma}\left(\alpha_s(\sqrt{s})\right) {\bar P} \exp \left[\int_{\sqrt{s}}^{{\sqrt{s}}/N}
\frac{d\mu}{\mu} \; \Gamma_{\! S \, ab \to tq\gamma}^{\dagger} \left(\alpha_s(\mu)\right)\right] \; \right.
\nonumber\\ && \left. \hspace{5mm} \times \,
{\tilde S}_{ab \to tq\gamma} \left(\alpha_s\left(\frac{\sqrt{s}}{N}\right)\right) \;
P \exp \left[\int_{\sqrt{s}}^{{\sqrt{s}}/N}
\frac{d\mu}{\mu}\; \Gamma_{\! S \, ab \to tq\gamma}
\left(\alpha_s(\mu)\right)\right] \right\} \, ,
\nonumber \\
\label{resummed}
\eeqa
where $P$ (${\bar P}$) denotes path-ordering in the same (reverse) sense as the integration variable $\mu$. This moment-space resummed cross section resums logarithms of the moment variable $N$.
The first exponential in Eq. (\ref{resummed}) resums soft and collinear emission from the initial-state partons, while the second exponential provides the scale evolution in terms of the parton anomalous dimensions $\gamma_{i/i}$. The third exponential describes radiation from the final-state quark. Explicit results for all these exponentials as well as for the soft anomalous dimensions $\Gamma_{\! S \, ab \to tq\gamma}$ (which are the same as those for $tqH$ production) at one and two loops can be found in Refs. \cite{FK2020,FK2021}. 

\subsection{Total cross sections at aNNLO}

The resummed cross section can be expanded to produce fixed-order expressions that can then be straightforwardly inverted to momentum space and used to calculate the soft-gluon corrections at higher orders (see e.g. Refs. \cite{NKsch,NKtW,NKtch,NKtW2016,NKNY,FK2020,FK2021,NKtt2l}). We denote the sum of the complete NLO cross section and the NNLO soft-gluon corrections as approximate NNLO (aNNLO).

\begin{table}[htbp]
\begin{center}
\begin{tabular}{|c|c|c|c|c|} \hline
\multicolumn{5}{|c|}{$tq\gamma$ cross sections at aNNLO for LHC energies with $p_{\gamma T}> 10$ GeV, $|\eta_{\gamma}| < 2.6$, and $\Delta R > 0.05$} \\ \hline
$\sigma$ in pb & 8 TeV & 13 TeV & 13.6 TeV & 14 TeV \\ \hline
MSHT20 (nnlo pdf) & $0.857^{+0.019}_{-0.020}{}^{+0.011}_{-0.007}$ & $2.30^{+0.05}_{-0.07} \pm 0.02$ & $2.50^{+0.05}_{-0.08} \pm 0.02$ & $2.65^{+0.06}_{-0.08} \pm 0.02$ \\ \hline
CT18 (nnlo pdf) & $0.864^{+0.018}_{-0.019}{}^{+0.022}_{-0.020}$ & $2.29^{+0.05}_{-0.07}{}^{+0.04}_{-0.05}$ & $2.51^{+0.06}_{-0.08}{}^{+0.04}_{-0.05}$ & $2.65^{+0.06}_{-0.08}{}^{+0.04}_{-0.05}$ \\ \hline
NNPDF4.0 (nnlo pdf) & $0.829^{+0.017}_{-0.018} \pm 0.003$ & $2.27^{+0.05}_{-0.07} \pm 0.01$ & $2.45^{+0.06}_{-0.08} \pm 0.01$ & $2.58^{+0.06}_{-0.08} \pm 0.01$ \\ \hline
\end{tabular}
\caption[]{The $tq\gamma$ cross sections at aNNLO (in pb, with central value for $\mu=m_t$ with scale and pdf uncertainties, and using cuts $p_{\gamma T}> 10$ GeV, $|\eta_{\gamma}| < 2.6$, $\Delta R > 0.05$) in $pp$ collisions with $\sqrt{S}=8$, 13, 13.6, and 14 TeV, with $m_t=172.5$ GeV and various NNLO pdf.}
\label{table7}
\end{center}
\end{table}

In Table 7 we display the aNNLO cross sections at 8, 13, 13.6, and 14 TeV energies using MSHT20, CT18, and NNPDF4.0 NNLO pdf sets, and the same cuts as in Tables 1 and 2. We see that we have significant increases at aNNLO relative to NLO of the order of 5\% to 6\%. 

\begin{table}[htbp]
\begin{center}
\begin{tabular}{|c|c|c|c|c|} \hline
\multicolumn{5}{|c|}{$tq\gamma$ cross sections at aNNLO for 13.6 TeV LHC energy with $|\eta_{\gamma}| < 2.6$ and $\Delta R > 0.05$ } \\ \hline
$\sigma$ in pb & $p_{\gamma T}> 10$ GeV  & $p_{\gamma T}> 15$ GeV & $p_{\gamma T}> 20$ GeV & $p_{\gamma T}> 25$ GeV \\ \hline
aNNLO (nnlo pdf) & $2.50^{+0.05}_{-0.08} \pm 0.02$ & $1.90^{+0.04}_{-0.06} \pm 0.02$ & $1.47^{+0.02}_{-0.04} \pm 0.01$ & $1.20^{+0.02}_{-0.03} \pm 0.01$     \\ \hline
\end{tabular}
\caption[]{The $tq\gamma$ cross sections at aNNLO (in pb, with central value for $\mu=m_t$ with scale and pdf uncertainties, and using cuts $p_{\gamma T}> 10$, 15, 20, and 25 GeV, as well as $|\eta_{\gamma}| < 2.6$ and $\Delta R > 0.05$) in $pp$ collisions with $\sqrt{S}=13.6$ TeV, with $m_t=172.5$ GeV and MSHT20 NNLO pdf.}
\label{table8}
\end{center}
\end{table}

Next, we investigate further the dependence of the cross section on kinematical cuts, as we did at LO and NLO in Section 2. 
In Table 8, we present the aNNLO cross sections with scale and pdf uncertainties (using MSHT20 NNLO pdf) at 13.6 TeV energy, with $|\eta_{\gamma}| < 2.6$ and $\Delta R > 0.05$, for four different  $p_{\gamma T}$ cuts (the same cuts as in Table 3). The values of the aNNLO cross sections diminish quickly with increasing cut on $p_T$, but the $K$-factors as well as the fractional scale and pdf uncertainties remain similar.

\begin{table}[htbp]
\begin{center}
\begin{tabular}{|c|c|c|c|c|} \hline
\multicolumn{5}{|c|}{$tq\gamma$ cross sections at aNNLO for 13.6 TeV LHC energy with $p_{\gamma T} > 10$ GeV and $\Delta R > 0.05$ } \\ \hline
$\sigma$ in pb & $|\eta_{\gamma}| < 1.5 $  & $|\eta_{\gamma}| < 2.0$ & $|\eta_{\gamma}| <  2.6$ & $|\eta_{\gamma}| < 3.0$ \\ \hline
aNNLO (nnlo pdf) & $1.55^{+0.03}_{-0.05} \pm 0.01 $ & $2.01^{+0.04}_{-0.06} \pm 0.02$ & $2.50^{+0.05}_{-0.08} \pm 0.02$ &  $2.77^{+0.06}_{-0.08} \pm 0.02 $  \\ \hline
\end{tabular}
\caption[]{The $tq\gamma$ cross sections at aNNLO (in pb, with central value for $\mu=m_t$ with scale and pdf uncertainties, and using cuts $|\eta_{\gamma}| < 1.5$, 2.0, 2.6, and 3.0, as well as $p_{\gamma T} > 10$ GeV and $\Delta R > 0.05$) in $pp$ collisions with $\sqrt{S}=13.6$ TeV, with $m_t=172.5$ GeV and MSHT20 NNLO pdf.}
\label{table9}
\end{center}
\end{table}

In Table 9, we present the aNNLO cross sections with scale and pdf uncertainties (using MSHT20 NNLO pdf) at 13.6 TeV energy, with $p_{\gamma T} > 10$ GeV and $\Delta R > 0.05$, for four different $|\eta_{\gamma}|$ cuts (the same cuts as in Table 4). The values of the aNNLO cross sections increase quickly with increasing upper bound on $|\eta_{\gamma}|$, but the $K$-factors and the fractional scale and pdf uncertainties remain similar.

\begin{table}[htbp]
\begin{center}
\begin{tabular}{|c|c|c|c|c|} \hline
\multicolumn{5}{|c|}{$tq\gamma$ cross sections at aNNLO for 13.6 TeV LHC energy with $p_{\gamma T} > 10$ GeV and $|\eta_{\gamma}| < 2.6$ } \\ \hline
$\sigma$ in pb & $\Delta R > 0.05$ & $\Delta R > 0.1$ & $\Delta R > 0.3$ & $\Delta R > 0.5$  \\ \hline
aNNLO (nnlo pdf) & $2.50^{+0.05}_{-0.08} \pm 0.02$ & $2.41^{+0.05}_{-0.07} \pm 0.02 $ & $2.23^{+0.04}_{-0.06} \pm 0.02 $ &  $ 2.10^{+0.04}_{-0.05} \pm 0.02 $  \\ \hline
\end{tabular}
\caption[]{The $tq\gamma$ cross sections at aNNLO (in pb, with central value for $\mu=m_t$ with scale and pdf uncertainties, and using cuts $\Delta R > 0.05$, 0.1, 0.3, and 0.5, as well as $p_{\gamma T} > 10$ GeV and $|\eta_{\gamma}| < 2.6$) in $pp$ collisions with $\sqrt{S}=13.6$ TeV, with $m_t=172.5$ GeV and MSHT20 NNLO pdf.}
\label{table10}
\end{center}
\end{table}

In Table 10, we present the aNNLO cross sections with scale and pdf uncertainties (using MSHT20 NNLO pdf) at 13.6 TeV energy, with $p_{\gamma T} > 10$ GeV and $|\eta_{\gamma}| < 2.6$, for four different $\Delta R$ cuts (the same cuts as in Table 5). The values of the aNNLO cross sections decrease with increasing cut on $\Delta R$, and the $K$-factor for the largest cut is significantly smaller than at the lower values for the cut, but the fractional scale and pdf uncertainties remain similar throughout.

For ${\bar t} q \gamma$ production with the same cuts as in Table 7 and using MSHT NNLO pdf, the aNNLO cross section is $1.61^{+0.04}_{-0.05}{}^{+0.02}_{-0.01}$ pb at 13 TeV, $1.77^{+0.05}_{-0.06}{}^{+0.02}_{-0.01}$ pb at 13.6 TeV, and $1.89^{+0.05}_{-0.07} \pm 0.02$ pb at 14 TeV.

Finally, we compare our aNNLO results for the sum of the $tq\gamma$ and ${\bar t} q \gamma$ cross sections with the data from the paper  with evidence of this process from CMS \cite{CMS13} as well as the paper with observation of this process from ATLAS \cite{ATLAS13}.

For the CMS cuts \cite{CMS13} of $p_{\gamma T}> 25$ GeV, $|\eta_{\gamma}| < 1.44$, $\Delta R > 0.5$ at 13 TeV, and using MSHT20 NNLO pdf, we find an aNNLO $tq \gamma$ cross section of $0.584^{+0.011}_{-0.015}{}^{+0.007}_{-0.005}$ pb and an aNNLO ${\bar t}q \gamma$ cross section of $0.406^{+0.010}_{-0.012}{}^{+0.005}_{-0.004}$ pb. Thus, the total $tq \gamma$+${\bar t}q \gamma$ cross section at aNNLO is $0.990^{+0.021}_{-0.027}{}^{+0.012}_{-0.008}$ pb which is a 5.8\% increase over the NLO result. Multiplying this total cross section by the branching fraction for $t \rightarrow \mu \nu b$ ($11.40 \pm 0.20 \%$), we find $(113 \pm 2)^{+2}_{-3} \pm 1$ fb. The measured value from CMS \cite{CMS13} is $115 \pm 17$ (stat.) $\pm 30$ (syst.) fb, but this includes further cuts on the decay products, and CMS compares their number to an NLO theory prediction with these extra cuts of $81 \pm 4$ fb. While we cannot directly apply the additional cuts on the decay products in our aNNLO prediction, we note that the 5.8\% enhancement at aNNLO (which would raise the central Standard Model prediction in \cite{CMS13} from 81 fb to 86 fb) considerably improves the comparison between theory and data.

For the ATLAS cuts \cite{ATLAS13} of $p_{\gamma T}> 20$ GeV, $|\eta_{\gamma}| < 2.37$, $\Delta R > 0.4$ at 13 TeV, and using MSHT20 NNLO pdf, we find an aNNLO $tq \gamma$ cross section of $1.115^{+0.018}_{-0.025}{}^{+0.011}_{-0.008}$ pb and an aNNLO ${\bar t}q \gamma$ cross section of $0.740^{+0.016}_{-0.021}{}^{+0.008}_{-0.007}$ pb. Thus, the total $tq \gamma$+${\bar t}q \gamma$ cross section at aNNLO is $1.855^{+0.034}_{-0.046}{}^{+0.019}_{-0.015}$ pb which is a 5.1\% increase over the NLO result. Multiplying this total cross section by the branching fraction for $t \rightarrow l \nu b$ ($33.2 \pm 1.0 \%$), we find $(616 \pm 19)^{+11}_{-15} {}^{+6}_{-5}$ fb. The measured value from ATLAS \cite{ATLAS13} is $580 \pm 19$ (stat.) $\pm 63$ (syst.) fb, but this includes further cuts on the decay products, and ATLAS compares their number to an NLO theory prediction with these extra cuts of $406^{+25}_{-32}$ fb. Again, we cannot directly apply the additional cuts on the decay products in our aNNLO prediction, but we note that the 5.1\% enhancement at aNNLO (which would raise the central Standard Model prediction in \cite{ATLAS13} from 406 fb to 427 fb) provides a considerable improvement in the comparison between theory and data.

\subsection{Top-quark $p_T$ and rapidity distributions at aNNLO}

In addition to the calculation of total cross sections, our resummation formalism allows the calculation of soft-gluon corrections to the top-quark differential distributions in transverse momentum and rapidity. 

\begin{figure}[htbp]
\begin{center}
\includegraphics[width=58mm]{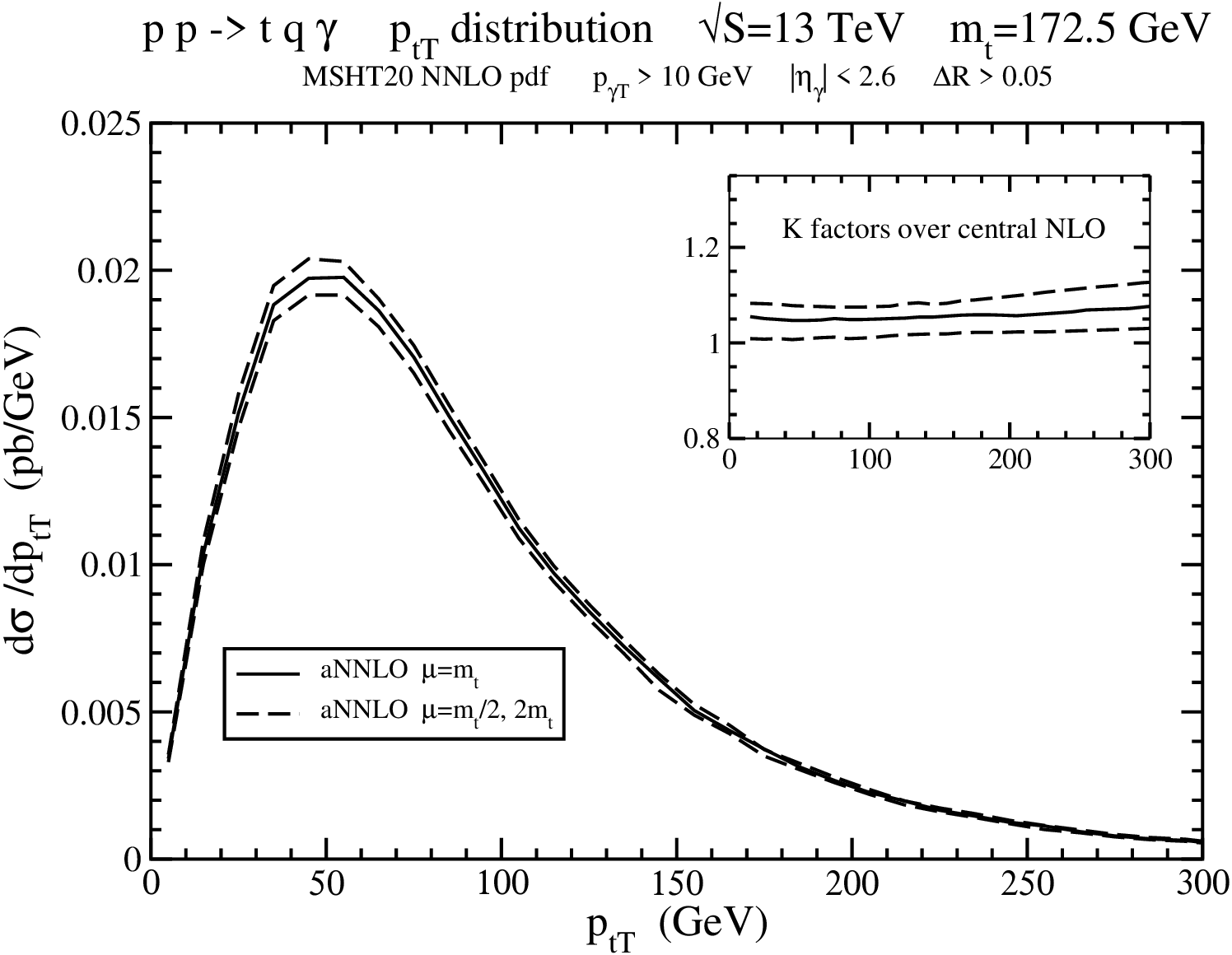}
\includegraphics[width=58mm]{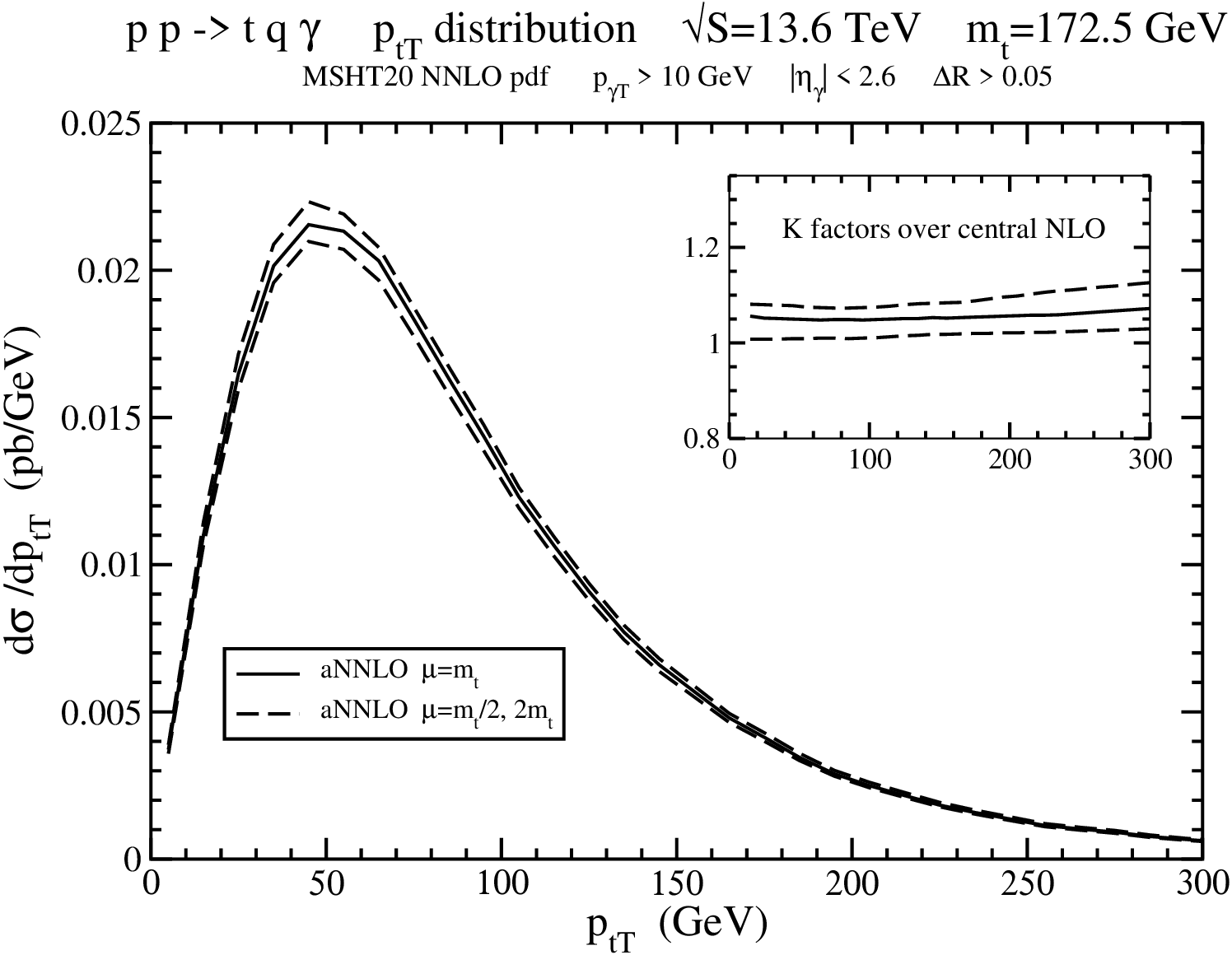}
\includegraphics[width=58mm]{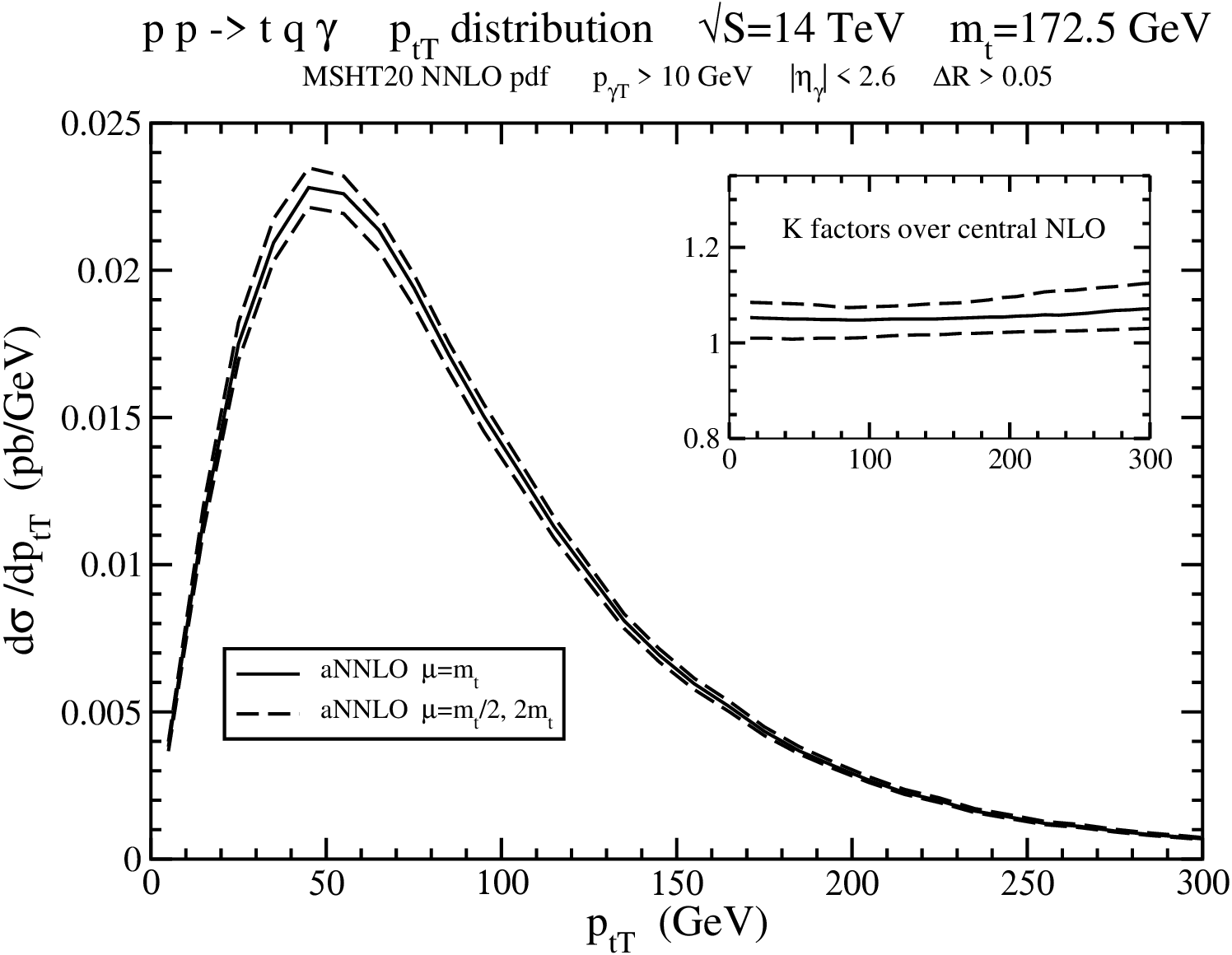}
\caption{The aNNLO top-quark $p_T$ distributions in $tq\gamma$ production at (left) 13 TeV, (middle) 13.6 TeV, and (right) 14 TeV LHC energies.}
\label{pTtqgamaNNLO}
\end{center}
\end{figure}

In Fig. \ref{pTtqgamaNNLO} we present the aNNLO top-quark $p_{tT}$ distributions, $d\sigma/dp_{tT}$, at 13 TeV (left), 13.6 TeV (middle), and 14 TeV (right) energies using MSHT20 NNLO pdf. We show the distributions for three different choices of scale. The inset plots show the $K$ factors with respect to the central ($\mu=m_t$) NLO $p_{tT}$ distribution. The enhancements at aNNLO relative to NLO are significant.

\begin{figure}[htbp]
\begin{center}
\includegraphics[width=57mm]{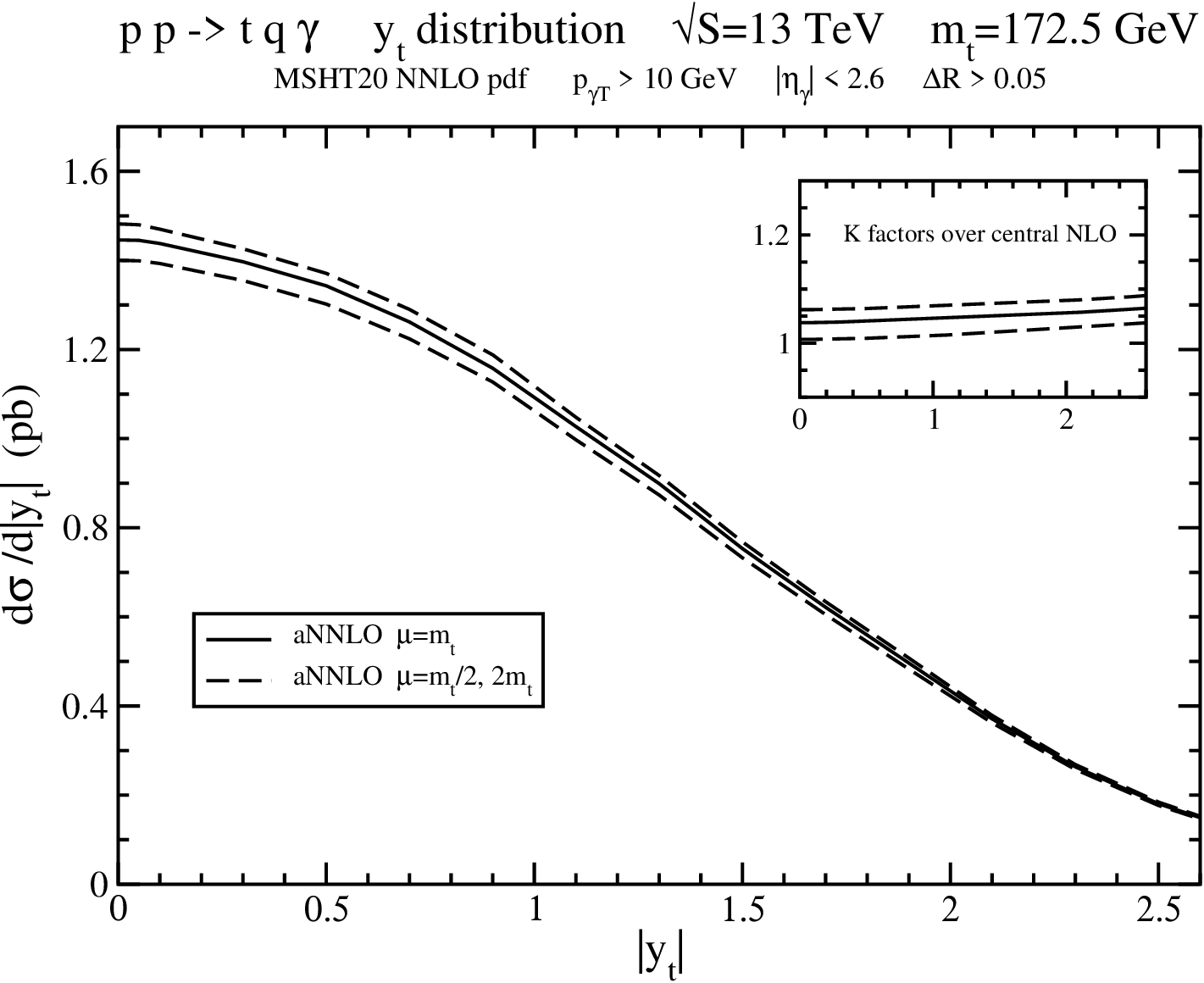}
\hspace{1mm}
\includegraphics[width=57mm]{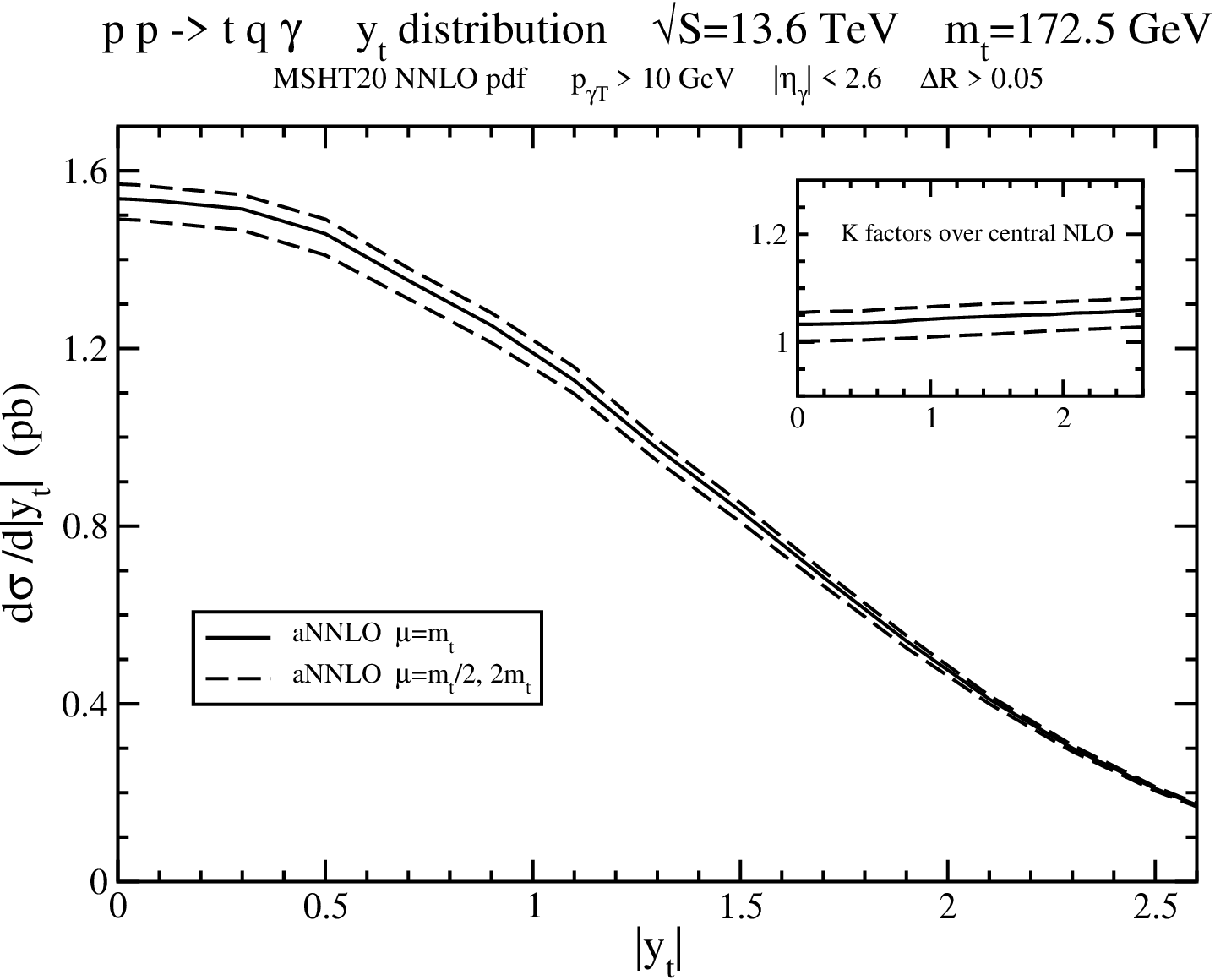}
\hspace{1mm}
\includegraphics[width=57mm]{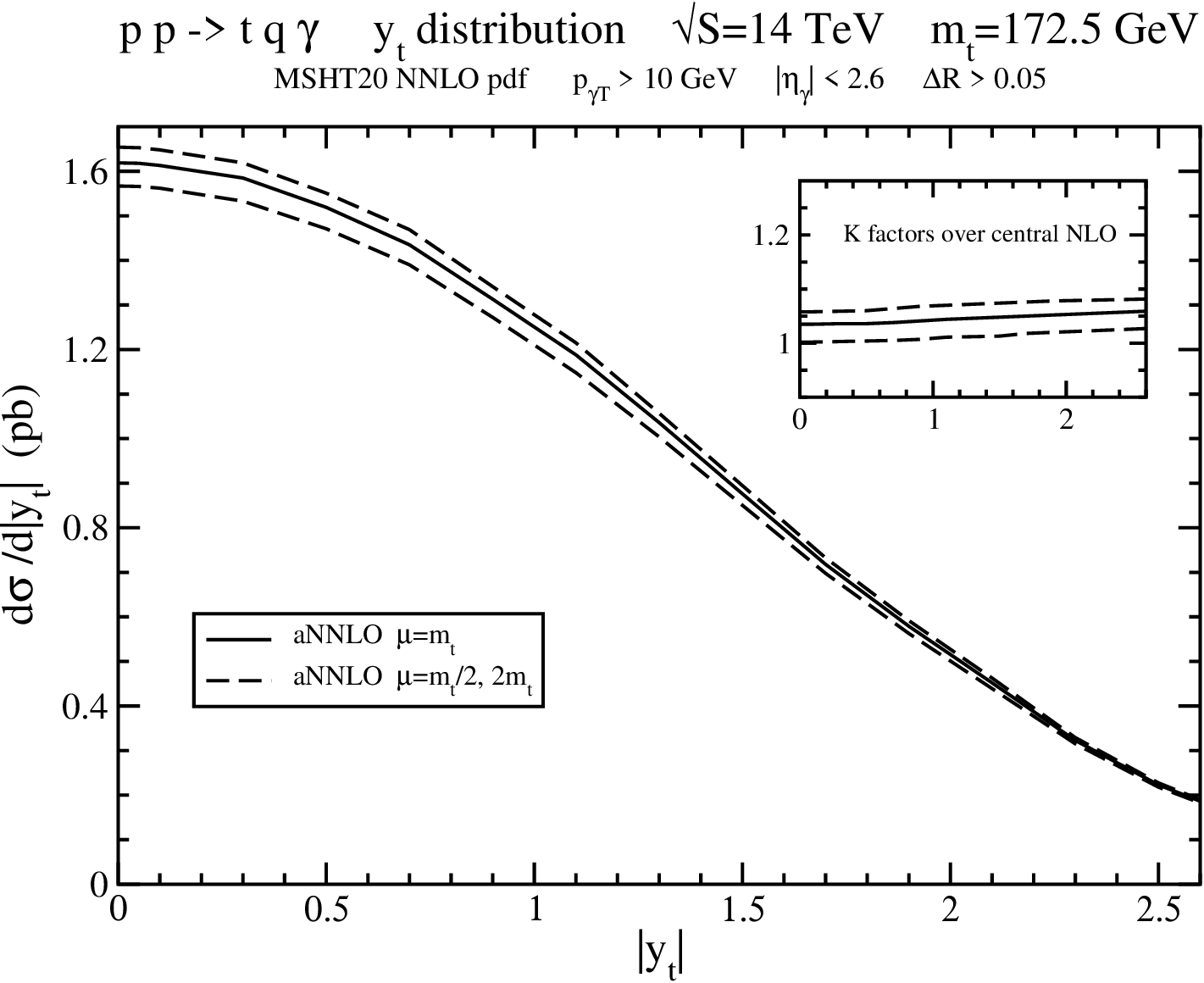}
\caption{The aNNLO top-quark rapidity distributions in $tq\gamma$ production at (left) 13 TeV, (middle) 13.6 TeV, and (right) 14 TeV LHC energies.}
\label{ytqgamaNNLO}
\end{center}
\end{figure}

In Fig. \ref{ytqgamaNNLO} we present the aNNLO top-quark rapidity distributions, $d\sigma/d|y_t|$, at 13 TeV (left), 13.6 TeV (middle), and 14 TeV (right) energies using MSHT20 NNLO pdf. Again, we show the distributions for three different choices of scale. The inset plots show the $K$ factors with respect to the central ($\mu=m_t$) NLO rapidity distribution and, again, we observe that the enhancements at aNNLO relative to NLO are significant.

\mysection{Conclusions}

We have presented a study of QCD corrections in $tq\gamma$ production at LHC energies as well as future collider energies up to 100 TeV. Precise theoretical estimates for this process are important since it is sensitive to anomalous electric and magnetic dipole moments as well as the top-quark charge and flavor-changing neutral currents.

We calculated the total cross sections at LO and NLO using a variety of kinematical cuts, and we investigated the scale dependence and the pdf uncertainties in the cross section. We also applied the theoretical formalism of soft-gluon resummation for $tq\gamma$ production in single-particle-inclusive kinematics in order to calculate aNNLO rates. The corrections at NLO are large and the further corrections at aNNLO are significant. The aNNLO enhancements of the total cross section predictions considerably improve the comparison of Standard Model theory with recent CMS \cite{CMS13} and ATLAS \cite{ATLAS13} measurements. 

We also calculated differential distributions for $tq\gamma$ production at LHC energies. In particular, we displayed results for the transverse-momentum and rapidity distributions of the top quark and the energy distribution of the photon. The QCD corrections at NLO provide large enhancements to all these distributions. We also calculated aNNLO corrections to the top-quark differential distributions and found them to provide further significant enhancements.

Our new aNNLO result provides a baseline of the Standard Model prediction, and is therefore definitely useful in the extraction of signals of new physics through anomalous moments of top quarks. Although the high-luminosity upgrade of the LHC may not able to reach the AMM and the EDM of the top quark expected from the scaling of the muon AMM deviation observed from the Standard Model prediction (as given in Ref. \cite{Fael:2013ira}), future collider experiments with higher energy and luminosity will certainly reach this projection.

\section*{Acknowledgements}
The work of N.K. is supported by the National Science Foundation under Grant No. PHY 2112025.

\end{document}